\documentclass[11pt, a4paper]{article}
\usepackage[margin=1.0in]{geometry}
\usepackage{authblk}
\usepackage{jcappub}
\usepackage{mathtools}
\usepackage{physics}
\usepackage{graphicx}					  
\graphicspath{{figs/}}

\usepackage{amssymb}
\usepackage{amsmath}
\usepackage{siunitx}
\usepackage{enumitem}
\usepackage{todonotes}
\usepackage{caption}
\usepackage{subcaption}
\usepackage{orcidlink}

\numberwithin{equation}{section}

\newcommand{\eq}{\text{eq}}

\newcommand{\const}{\text{const}}

\newcommand{\pl}{\text{Pl}}
\newcommand{\mpl}{m_\text{Pl}}
\newcommand{\fNL}{f_\text{NL}}
\newcommand{\lrfrac}[2]{\left( \frac{#1}{#2} \right)} 
\newcommand{\phicmb}{\phi_\text{CMB}}

\DeclareSIUnit\year{yr}
\DeclareSIUnit\parsec{pc}
\DeclareSIUnit\Ms{M_{\odot}}

\usepackage[splitrule]{footmisc}

\title{Ultra-slow-roll Inflation with Non-perturbative Non-Gaussianity and Scalar Induced Gravitational Waves}
\date{\today}

\author[a]{Manuel Drees\orcidlink{0000-0001-5579-7780}}
\author[a]{Chenhuan Wang\orcidlink{0009-0005-0858-050X}}

\emailAdd{mandrees@uni-bonn.de}
\emailAdd{cwang1@uni-bonn.de}

\affiliation[a]{Bethe Center for Theoretical Physics and Physikalisches
  Institut, Universit{\"a}t Bonn, \\ Nussallee 12, 53115 Bonn, Germany}

\abstract{An ultra-slow-roll phase during inflation could potentially
  produce the right abundance of primordial black holes (PBHs) for
  them to form all of dark matter (DM). We consider such a scenario,
  carefully treating the transitions from slow to ultra slow roll
  inflation and back, using a parametrisation of the inflaton
  potential that can describe inflation from the time when CMB scales
  crossed out of the horizon until its end. A $\delta N$ analysis
  shows that this model can possess $\order{1}$ non-Gaussianity. When
  computing the primordial black hole abundance, we keep the full
  non-linear relation between curvature perturbations and the density
  contrast and consider the non-Gaussianity to all orders. We find
  that $\order{1}$ non-Gaussianity is sufficient to enhance the PBH
  abundance by orders of magnitude, leading to reduced gravitational
  wave (GW) signals for fixed PBH abundance. The signal to noise ratio
  is computed for future gravitational wave observatories. We find
  that, in spite of the reduced strength, the GW signal is well above
  the sensitivity of LISA if PBHs form a significant component of DM.}

\begin{document}
\maketitle

%%%%%%%%%%%%%%%%%%%%%%%%%%%%%%%%%%%%%%%%%%%%%%%%%%%%%%%%%%%%%%%%%%%%%%%%%%%%%%%

\section{Introduction}

Inflation is the most promising extension to the (old) standard
cosmology \cite{planck10, starobinskyNewTypeIsotropic1980}. It is
defined by a period of exponential expansion of the Universe,
presumably driven by an inflaton field
\cite{guthInflationaryUniversePossible1981}. In addition to solving a
number of open problems in standard cosmology, inflation provides a
mechanism to generate the seeds of the temperature anisotropies
observed in the cosmic microwave background (CMB) via quantum
fluctuations of the inflaton field leading to density perturbations;
the same perturbations also act as seeds of large-scale structures in
the late Universe \cite{mukhanovQuantumFluctuationsNonsingular1981,
  hideokodamaCosmologicalPerturbationTheory1984,
  baumannTASILecturesInflation2012}.

The simplest implementation of this idea is single-field slow-roll
inflation, where a (real) scalar field rolls down its potential slowly
and sources the exponential expansion of the Universe. The
anisotropies observed in the CMB put constraints on the primordial
perturbations, but only at length scales around
$\sim \SI{0.01}{\mega\parsec}$ \cite{planck10}; these lead to
constraints on the inflaton potential for some range of values of the
inflaton field, typically about $50$ $e$-folds of expansion before the
end of inflation. Moreover, in order to solve the horizon problem, at
least $60$ or so $e$-folds of total inflation are required
\cite{kolb-turner, gorbunovIntroductionTheoryEarly2011,
  baumannTASILecturesInflation2012}. However, the primordial
fluctuations at smaller length scales, and the inflaton potential
during the last $\sim 40$ $e$-folds of inflation, are largely
unconstrained.

Since the first direct detection of gravitational waves (GW) by LIGO
\cite{collaborationImprovedAnalysisGW1509142016}, the possible
existence of primordial black holes (PBHs) has gained renewed
interests. One of the production mechanism is by the collapse of
primordial fluctuations, which would need to be several orders of
magnitude larger than those that seed the CMB anisotropies
\cite{escrivaPrimordialBlackHoles2022,
  carrPrimordialBlackHoles2020}. One appeal of PBHs is that they might
be produced with a wide range of masses, including values that are not
easily explained through stellar collapse; this explains their
relevance for the interpretation of LIGO data. In fact, PBHs could
well be of sub-solar mass. For masses in the ``asteroid'' range,
around $10^{-15} M_\odot$, they can even account for the entirety of
(cold) dark matter \cite{carrPrimordialBlackHoles2020,
  carrConstraintsPrimordialBlack2021}.

As already noted, PBH formation requires large density fluctuations.
Restricting ourselves to single-field inflation, sufficiently large
primordial perturbations can be produced during a so-called
ultra-slow-roll (USR) phase: a very flat (or even very slightly
rising) part of the inflaton potential slows down the inflaton field
dramatically via Hubble friction, causing the power spectrum of
primordial density perturbations to grow by several orders of
magnitude. Additionally, such evolution can introduce up to
$\order{1}$ non-Gaussianity (NG) in the probability distribution of
the fluctuations, which generally boosts the PBH abundance. This
scenario has gained a lot of attention lately
\cite{firouzjahiInducedGravitationalWaves2023, caiOneSmallStep2022,
  caiRevisitingNonGaussianityNonattractor2018,
  frosinaInflationaryInterpretationNHz2023}. In this work, we use a
perfectly flat plateau in the inflaton potential to model the USR
phase, similar to
ref.~\cite{caiRevisitingNonGaussianityNonattractor2018}; however, we
describe the post-USR phase of inflation with a cubic potential, which
allows a consistent description of inflation until it ends. The
location of this plateau essentially fixes the range of masses of the
PBHs, while their abundance depends on the width of the plateau as
well as on the ``smoothness parameter'' describing the transition from
the USR phase to the second slow-roll phase.

It is not at all trivial to predict the PBH abundance from the
primordial fluctuations. In the past, the simple Press-Schechter
formalism with the approximation that the density contrast is linearly
in the curvature perturbation, $\delta\rho \propto \zeta$, was often
used. Recently, it has been advocated in
refs.~\cite{escrivaPrimordialBlackHoles2022,
  gowNonperturbativeNonGaussianityPrimordial2022,
  delucaNoteAbundancePrimordial2022,
  muscoThresholdPrimordialBlack2019,
  ferrantePrimordialNongaussianityAll2023} that these approximations
are not sufficient, especially when sizable non-Gaussianities are
present. In this work, we adopt their method of calculating the PBH
abundance, while using the $\delta N$ formalism to calculate the
amount of local non-Gaussian probability distribution of the
primordial fluctuations.  \cite{sasakiGeneralAnalyticFormula1996,
  lythGeneralProofConservation2005}.

Cosmological perturbations with different helicities are decoupled at
first order \cite{domenechScalarInducedGravitational2021,
  gorbunovIntroductionTheoryEarly2011}. At second order, however,
tensor modes can be produced by scalar curvature perturbations. As PBH
formation generically requires large primordial fluctuations, it is
expected that gravitational waves (GW) are generated at second
order. As we shall see, for asteroid-mass PBHs the accompanying GWs
have frequencies of $\order{0.1}\si{\hertz}$. We find that if PBHs
in this mass range contribute significantly to the total dark matter
(DM) density, the associated primordial GWs would have a strength well
above the sensitivity of future space based GW observatories.

The remainder of this work is organized as follows. We introduce the
basic equations and our inflation model in section~\ref{sec:setup}. The
power spectrum is computed and discussed in section~\ref{sec:PS}. In
section~\ref{sec:delta_N}, a $\delta N$ calculation is carried out to
compute the non-Gaussian probability distribution of the curvature
perturbations. Then the methods for computing PBH abundance and
gravitational waves are explained in section~\ref{sec:PBH}. We present
our results in section~\ref{sec:result}, while section~\ref{sec:conclusion}
contains our conclusion and some discussions.

We acknowledge the use of the following computational tools:
\cite{harris2020array, 2020SciPy-NMeth, Hunter:2007,
  witkowskiSIGWfastPythonPackage2022}. We use natural units where
$c=\hbar =k_B = 1$ and use the reduced Planck mass
$\mpl = (8\pi G)^{-1/2} \simeq \SI{2.4e18}{\giga\eV}$.

%%%%%%%%%%%%%%%%%%%%%%%%%%%%%%%%%%%%%%%%%%%%%%%%%%%%%%%%%%%%%%%%%%%%%%%%%%%%%%%

\section{Setup}\label{sec:setup}

In this section, we introduce the basic equations that describe the
dynamics of inflaton. Several important quantities are defined as
well.

We start with the Friedmann-Lemaître-Robertson-Walker (FLRW) metric,
which describes the homogeneous and isotropic background spacetime by
the line element
\begin{equation} \label{e2.1}
    \dd{s}^2 = \dd{t}^2 - a^2(t) \dd{\vec{x}}^2\,;
\end{equation}
here $a(t)$ is the scale factor which characterizes the cosmic
expansion and $t$ is the cosmic time. We introduce a single, minimally
coupled scalar inflaton field $\phi(t)$; as usual we assume that it is
homogeneous, i.e. depends only on $t$. The evolution of the
background is then governed by the equation of motion for $\phi$
and the Friedmann equations:
\begin{subequations}
\label{eq:backgrounds}
\begin{align}
& \ddot{\phi} + 3 H \dot{\phi} + V'(\phi) = 0\,, \label{eq:homo-eom} \\
& H^2 = \frac{1}{3 m_\pl^2} \left( \frac{1}{2} \dot{\phi}^2 + V(\phi)
    \right)\, ,  \label{eq:1-fried}\\
& \dot{H} = - \frac{\dot{\phi}^2}{2 \mpl^2}\,. \label{eq:2-fried}
\end{align}
\end{subequations}
Here the Hubble parameter is defined as $H = \dot{a}(t)/a(t)$, the
overdot denoting a derivative with respect to $t$; a prime is used for
the derivative of the potential,
$V'(\phi) = \dd{V(\phi)} / \dd{\phi}$, $V(\phi)$ being the inflaton
potential which will be specified below.

The most important inflationary observables can be computed from the
potential slow-roll parameters, defined as
\cite{baumannTASILecturesInflation2012}
\begin{equation}     \label{eq:pot-SR-para}
  \epsilon_V(\phi) = \frac{\mpl^2}{2 } \left( \frac{V'(\phi)}{V(\phi)} \right)^2
  \,, \quad \eta_V(\phi) = \mpl^2 \frac{V''(\phi)}{V(\phi)}\,.
\end{equation}
Slow-roll inflation occurs if the first derivative of the inflaton
potential is small, $\epsilon_V \ll 1$; $|\eta_V| \ll 1$ is required
in addition if inflation is to last more than a couple of
$e$-folds (defined to be $N(t) = \ln a(t)$). 
On the other hand, the Hubble slow-roll parameters~\cite{planck10} characterize the rate of change of the Hubble
parameter:\footnote{Note that another Hubble slow-roll parameter
  (often written as $\eta_H$) is sometimes used:
  $\eta_H = - \ddot{H}/(2H \dot{H}) = - \ddot{\phi}/(H \dot{\phi})$.}
\begin{equation} 	\label{math:Hubble-SR}
  \epsilon_1 = - \frac{\dot{H}}{H^2}\,, \quad
  \epsilon_2 = \frac{\dot{\epsilon_1}}{H \epsilon_1}\,.
\end{equation}
For single field inflation described by eqs.~(\ref{eq:backgrounds}) these
become:
\begin{equation} \label{eq:epphi}
\epsilon_1 = \frac{1}{2\mpl^2} \frac{\dot{\phi}^2}{H^2}\,, \quad
\epsilon_2 = \frac{\ddot{H}}{\dot{H}} \frac{1}{H} - 2 \frac{\dot{H}}{H^2}
= 2 \frac{\ddot{\phi}}{\dot{\phi}H} + 2 \epsilon_1\,.
\end{equation}
During slow-roll inflation the $\ddot{\phi}$ term in
eq.~(\ref{eq:homo-eom}) can be neglected, so that
$\dot\phi \simeq - V'/(3H)$, with $H^2 \simeq V/(3\mpl^2)$. Hence, to
good approximation $\epsilon_1 \simeq \epsilon_V$ during slow-roll
\cite{navasReviewParticlePhysics2024}. The slow-roll condition
$\epsilon_V \ll 1$ therefore implies that $H$ changes slowly during
inflation.

In this work, we are primarily interested in a so-called
ultra-slow-roll (USR) phase, where the gradient of the potential
(almost) vanishes for a small section of the inflaton potential. In
our simplified, phenomenological model, the potential is taken to be
exactly flat for some range of $\phi$. The Universe is still
inflating, but the field is decelerating (exponentially, as we will
see shortly). Setting $V' = 0$ in eq.~\eqref{eq:homo-eom} implies
$\epsilon_2 \rightarrow -6$ during USR, while the deceleration of the
field implies that $\epsilon_1$ is even smaller than in the preceding
slow-roll phase.

We aim to model the inflation potential from CMB scales to the end of
inflation, with a USR phase somewhere in between. For defineteness we
take $\phi$ to initially be positive, with $\phi = 0$ defining the
true (post-inflationary) minimum; this means that $\phi$ decreases
during inflation. The USR phase then occurs for
$\phi \in [\phi_e, \phi_i]$. Prior to the USR phase ($\phi > \phi_i$),
terms up to quadratic order in the potentials are sufficient to
describe the dynamics up to CMB scales, so that the coefficients of
the polynomial can be conveniently expressed using the potential
slow-roll parameters at $\phi_i$. We want our ansatz for the post-USR
phase, i.e. for $\phi < \phi_e$, to allow for a smooth transition,
i.e. $V'(\phi_e) = 0$, while also describing the minimum of the
potential at $\phi = 0$. The following potential satisfies these
requirements:
\begin{equation}    \label{eq:inf-pot-new}
   V(\phi) = V_0 \cdot 
  \begin{cases}
    1 + \sqrt{2\tilde{\epsilon}_V} \frac{\phi - \phi_i}{\mpl}
    + \frac{\tilde{\eta}_V}{2} \lrfrac{\phi - \phi_i}{\mpl}^2\,,
    & {\rm for} \ \phi > \phi_i\,,  \\
    1\,, & {\rm for} \ \phi_e \leq \phi \leq \phi_i\,, \\
    \frac{3}{1+2\beta} \lrfrac{\phi}{\phi_e}^2 \left[ 1
      - \frac{2}{3} \frac{\phi}{\phi_e} (1-\beta) \right]\,,
    & {\rm for} \ \phi < \phi_e\,.
    \end{cases}
\end{equation}
$\tilde{\epsilon}_V,\; \tilde{\eta}_V,\; \phi_i,\; \phi_e, \;V_0$ and
$\beta$ are the free parameters of our potential. The values of these
constants remain to be determined. The potential is continuous at
$\phi_i$ and $\phi_e$, and the smoothness of its first derivative at
$\phi_e$ is controlled by the parameter $\beta$.

The six free parameters defining the potential~\eqref{eq:inf-pot-new},
as well as the field value at the CMB scale $\phicmb$ define our model
completely. It is required to reproduce the observables at the CMB
scale: the normalization of the power spectrum,
$\mathcal{A}_\mathcal{R} = \num{2.10e-9}$~\cite{planck10}, and the
spectral index $n_s = 0.974$~\cite{louisAtacamaCosmologyTelescope2025}. This leaves us with five
free parameters. We keep $\tilde{\epsilon}_V$ as a free parameter,
but trade other parameters for the number of $e$-folds of inflation
that occur in the three epochs defined by eq.~(\ref{eq:inf-pot-new}).
In particular, $N_+$ denotes the number of $e$-folds between CMB scale
and USR; since this occurs during slow-roll, we have
\cite{gorbunovIntroductionTheoryEarly2011}
\begin{equation} \label{eq:npl}
  N_+ (\phi_\text{CMB} - \phi_i, \eta_V) \simeq \int_{\phi_i}^{\phicmb}
  \dd{\phi} \frac{V(\phi)}{V'(\phi)}\,.
\end{equation}
Moreover, the spectral index at the CMB scale is related to the
potential slow-roll parameter at $\phi_\text{CMB}$
\cite{navasReviewParticlePhysics2024,
  gorbunovIntroductionTheoryEarly2011,
  baumannTASILecturesInflation2012}:
\begin{equation} \label{eq:ns}
    n_s = 1 - 6\epsilon_V(\phi_\text{CMB}) + 2 \eta_V (\phi_\text{CMB})\,.
\end{equation}
Note that the potential slow-roll parameters defined in
eq.~\eqref{eq:pot-SR-para} at the CMB scale, $\epsilon_V(\phicmb)$ and
$\eta_V(\phicmb)$, can be quite different from the constants
$\tilde{\epsilon}_V$ and $\tilde{\eta}_V$ if $\phicmb - \phi_i$ is not
much smaller than $\phi_i$. We use eqs.~(\ref{eq:npl}) and (\ref{eq:ns})
to determine $\phicmb - \phi_i$ and $\tilde\eta_V$.

The overall scale $V_0$ of the inflaton potential is fixed by the
normalization of the power spectrum $\mathcal{A}_\mathcal{R}$ at the
CMB scale, once $\phicmb - \phi_i$ is known. The difference
$\phi_e - \phi_i$ can be computed from the desired duration of USR
phase $\Delta N$.\footnote{The numerical determination of $\Delta N$
  requires careful treatment. The initial velocity of the inflaton
  field $\dot{\phi}$ has to be computed numerically using the
  potential at $\phi > \phi_i$. $\phi_e - \phi_i$ is then determined from
  $\Delta N$ by integrating the inflaton equation of motion.}

Finally, $\beta$ and $\phi_e$ determine the duration of the second
slow-roll phase $N_-$. In practice, we fix the total number of
$e$-folds of {\em slow-roll inflation} after the CMB scale has exited
the horizon to be $N_\text{tot} - \Delta N=55$, where the exact
numerical value is not very important for our main results. We do this
mainly for technical reasons. Fixing the total number of $e$-folds of
inflation would cause the duration of the first SR phase, $N_+$, to
depend on $\Delta N$. The final PBH DM fraction $f_\text{PBH}$ would
then not be a simple monotonic function of $\Delta N$. Hence, there are
only four free parameters left in the end:
$\tilde{\epsilon}_V,\; \beta,\; \Delta N$ and $N_-$.

An important observational constraint is given by the upper bound on
the tensor-to-scalar ratio bound at the CMB scale,
$r = 16\epsilon_{V}(\phicmb) < 0.056$~\cite{planck10}. The consistency
of any chosen set of parameters with the $r$-bound has to be checked.
Moreover, theoretical consistency of our treatment imposes an upper
bound on the duration $\Delta N$ of the USR phase. The classical field
variation in a Hubble time $\Delta t \simeq H^{-1}$ is
$\Delta \phi \simeq \dot{\phi}/H$, whereas quantum fluctuations are of
order $\delta \phi \sim H/(2\pi)$. If the quantum fluctuations
dominate over the classical evolution, $\dot{\phi}/H^2 \lesssim 1$, a
deterministic evolution of the inflaton field is not possible
anymore. This would lead to eternal inflation~\cite{gorbunovIntroductionTheoryEarly2011,
  guthEternalInflationIts2007}. We find this generically happens for
$\Delta N \gtrsim 3$. However, such a large value of $\Delta N$ almost
always leads to PBH overproduction anyway.

With the potential defined by eq.~\eqref{eq:inf-pot-new}, the potential
slow-roll parameters at the beginning of the second slow-roll phase,
$\phi = \phi_e$, are:
\begin{equation}  \label{eq:pot-SR-para-real}
  \epsilon_V(\phi_e) = 18 \left( \frac{\mpl}{\phi_e} \frac{\beta}{1+2\beta}
  \right)^2\,, \quad
  \eta_V(\phi_e) = 6 \lrfrac{\mpl}{\phi_e}^2 \frac{2\beta - 1}{1 + 2\beta}\,.
\end{equation}
Thus, an extended period of slow-roll inflation after USR
($\phi < \phi_e$) is only possible for $\phi_e/\mpl$ (much) larger
than unity.\footnote{This conclusion could be avoided by using a
  quartic polynomial to describe the potential at $\phi < \phi_e$, as
  in ref.~\cite{Drees:2021wgd}. However, our ansatz is already
  sufficiently general to allow us to vary $N_-$ and
  $\epsilon_V(\phi_e)$ in wide ranges while obeying the bound on $r$,
  without having to introduce one more free parameter.}  Inflation
ends at $\phi = \phi_{\rm end} \simeq \sqrt{2} \mpl$ where
$\epsilon_V \sim \eta_V \sim 1$. The number of $e$-folds after the end
of the USR phase, $N_-$, can be computed from eq.~(\ref{eq:npl}) with
$\phi_i \rightarrow \phi_{\rm end}$ and
$\phi_{\rm CMB} \rightarrow \phi_e$. We are interested in scenarios
with $N_- \sim 20$, so that (large) fluctuations that cross out of the
horizon around the USR phase generate asteroid mass PBHs (see section \ref{sec:pbhm}). In this case
$\phi_e/\mpl$ cannot be made too large, regardless of the choice of
$\beta$, since otherwise $N_-$ would become too large. Hence, the
second potential slow-roll parameter $\eta_V$ has typical magnitude
$\order{0.1}$ during the second slow-roll epoch.

A crucial quantity determining the size of curvature perturbations and
non-Gaussianity is the ``smoothness parameter'' $h$, where $h = 0$
means that not only the value of the potential, but also its first
derivative is smooth at $\phi = \phi_e$; the first
eq.~(\ref{eq:pot-SR-para-real}) shows that $h = 0$ for $\beta = 0$. $h$
is defined via the ``velocity'' of the inflaton field in $N$-space,
$\pi = \dd\phi / \dd N = \dot{\phi} / H(\phi)$; more exactly, it is
six times the ratio of the absolute value of the slow-roll prediction
of field velocity at the beginning of the second SR phase,
$|\pi_0| = \mpl \sqrt{2\epsilon_V(\phi_e)}$, to the actual (numerical)
velocity at that point, $\pi_e$. Using eq.~\eqref{eq:pot-SR-para-real}
for $\epsilon_V(\phi_e)$, and eq.~(\ref{math:USR_phi}) from
section~\ref{sec:delta_N} with $\pi_i = \mpl \sqrt{2 \tilde\epsilon_V}$
for the velocity at the beginning of the USR phase, we have
\begin{equation}   \label{eq:smooth-h}
  h \equiv 6 \frac{\mpl \sqrt{2\epsilon_V(\phi_e)}} {\pi_e}
  = -18 \sqrt{ \frac{2} {\tilde{\epsilon}_V} }
  \frac {\mpl} {\phi_e} \frac{\beta}{1+2\beta} e^{3\Delta N}\,.
\end{equation}
Hence, for a smooth transition, $|h| \ll 1$, the inflaton field has to
gradually slow down to reach the SR attractor; since in this scenario
$\epsilon_V$ increases relatively quickly for $\phi < \phi_e$,
``catching up'' with the slow-roll trajectory can take a few
$e$-folds. On the other hand for $|h| \gg 1$, $\epsilon_V$ makes a
sizable jump at $\phi = \phi_e$ but then remains essentially constant,
leading to a faster transition.

\section{Curvature Perturbation}
\label{sec:PS}

Once the inflaton potential is specified as in
eq.~(\ref{eq:inf-pot-new}), the dynamical evolution of the background
can easily be obtained by integrating eqs.~(\ref{eq:backgrounds}). Here
we are more interested in the spectrum of quantum fluctuations. As
usual, during inflation they are stretched to super-horizon size. In
this section we briefly review the calculation of the curvature
perturbations and their power spectrum created during
inflation.

In the conformal Newtonian gauge without anisotropic stress, the FLRW
metric with scalar perturbation $\Phi$ can be expressed as
\cite{gorbunovIntroductionTheoryEarly2011,
  navasReviewParticlePhysics2024}
\begin{equation}
  \dd{s}^2 = a^2(\eta) \left[ (1+2\Phi) \dd{\eta}^2
    - (1-2\Phi)\dd{\vec{x}}^2 \right]\,,
\end{equation}
where the conformal time $\eta \equiv \int \dd{t}/a(t)$ is used. The
inflaton field fluctuates around its classical value $\phi(t)$ by
$\delta \phi(\vec{x}, t)$. The gauge invariant spatial curvature
$\mathcal{R}$ can be defined as
\cite{gorbunovIntroductionTheoryEarly2011,
  navasReviewParticlePhysics2024}
\begin{equation}
  \mathcal{R} = - \Phi - \frac{\partial_\eta a}{a}
  \frac{\delta \phi}{\partial_\eta \phi}\,.
\end{equation}
It coincides with the curvature perturbation $\zeta$ in slow-roll
inflation, and more generally on superhorizon scales
\cite{baumannTASILecturesInflation2012}. As already noted, $\zeta$ and
$\mathcal{R}$ are conserved on superhorizon scales, $k\ll aH$, during
slow-roll inflation.

Primordial black holes can only form from the collapse of regions with
large overdensity, which requires large curvature fluctuations. The
properties of a Gaussian random field are fully determined by the
power spectrum or two-point correlation function. In order to compute
this quantity in a general setting, allowing in particular for a USR
phase, we define the Mukhanov variable $u \equiv z \mathcal{R}$ and
$z \equiv a\dot{\phi}/H$. After promoting $u$ to a quantum field, its
mode function in Fourier space\footnote{A quantum fluctuation is
  characterized by a wave vector $\vec k$. However, in an isotropic
  background the dynamics only depends on $k = |\vec k|$, and one can
  write $u_{\vec k} = u_k \eta(\vec{k}/k)$ with $|\eta| = 1$.} $u_k$
follows the Mukhanov-Sasaki equation
\cite{gorbunovIntroductionTheoryEarly2011,
  baumannTASILecturesInflation2012, navasReviewParticlePhysics2024}
\begin{equation}    \label{eq:ms-eq}
\partial_\eta^2 u_k + \left(k^2 - \frac{\partial_\eta^2 z}{z} \right) u_k = 0.
\end{equation}
At sufficiently early times (formally in the infinite past), the
fluctuations were well inside the cosmic horizon. In that case they
behave just like in flat spacetime, since $1/k \ll 1/(aH)$, and the
so-called Bunch-Davies vacuum can be used as initial conditions
\cite{baumannTASILecturesInflation2012,
  n.d.birrellQuantumFieldsCurved1984}:
\begin{equation}    \label{eq:bunch-davis}
  \lim_{k \gg aH} \partial_\eta u_k = -ik u_k, \quad \lim_{k \gg aH}|u_k|
  = \frac{1}{\sqrt{2k}}\,.
\end{equation}
The primordial power spectrum for curvature perturbations
$\mathcal{P}_\mathcal{R}$, defined via the expectation value in
Fourier space
\begin{equation} \label{eq:power_def}
  \expval{\mathcal{R}_{\vec{k}} \mathcal{R}_{\vec{k}'}} = (2\pi)^3
  \delta^{(3)} \left(\vec{k} + \vec{k}'\right)
  \frac{2\pi^2}{k^3} \mathcal{P}_\mathcal{R}(k)\,,
\end{equation}
can easily be computed from $u_k$ as
\begin{equation}     \label{eq:PS-u}
\mathcal{P}_{\mathcal{R}} = \frac{k^3}{2\pi^2} \left| \frac{u_k}{z} \right|^2\,;
\end{equation}
the right-hand side of eq.~(\ref{eq:PS-u}) should be 
evaluated at the time when the mode $k$ is well outside the horizon $k \ll aH$.
% computed at the end of inflation. 
In the usual slow-roll inflation, perturbations get
frozen after horizon crossing, and the power spectrum can be directly
computed from the inflaton potential
\cite{baumannTASILecturesInflation2012,
  navasReviewParticlePhysics2024}:
\begin{equation} \label{eq:PS-SR}
    \mathcal{P}_\mathcal{R} = \frac{1}{2\pi^2} \lrfrac{H^4}{\dot{\phi}^2}_*
    = \frac{1}{2\pi^2} \lrfrac{H^2}{2\mpl^2 \epsilon_V}_*\,;
\end{equation}
here $H$ and $\dot{\phi}$ should be evaluated when the perturbation
with comoving wave vector $\vec k$ crossed out of the horizon,
$|\vec k| = a_* H_*$.

In order to see how a perturbation evolves during USR, we need to go
back to eq.~\eqref{eq:ms-eq}. At superhorizon scales, i.e. for
$k \ll a H$, the spatial curvature can be written as\footnote{This can
  be most easily obtained by writing $z = \sqrt{2\epsilon_V} a \mpl $
  via eq.~\eqref{eq:2-fried}.} \cite{byrnesSteepestGrowthPower2019,
  piPrimordialBlackHole2023}
\begin{equation}    \label{eq:R-solution}
 \mathcal{R}_k \simeq A_k + B_k \int^t \frac{\dd t'}{a^2(t') \epsilon_V(t')}\,,
\end{equation}
with two integration constants $A_k$ and $B_k$. During the usual
slow-roll inflation, the curvature perturbation has a constant and a
decaying solution, as $\epsilon_V \approx \text{const}$. Hence, it is
conserved on superhorizon scale. In USR, however, $\epsilon_V$
approaches zero and the decaying mode becomes the growing mode
\cite{byrnesSteepestGrowthPower2019, piPrimordialBlackHole2023} (see
also ref.\cite{dreesOvershootingCriticalHiggs2021} for a similar
argument). Hence, curvature perturbations may no longer be conserved
at super horizon scales, but can grow during USR.

\begin{figure}[ht]
    \centering
    \includegraphics[width=0.48\textwidth]{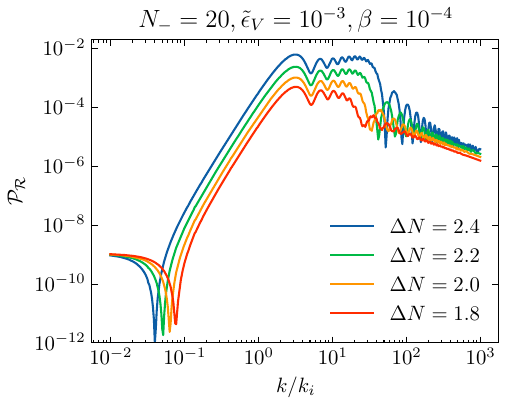}
    \includegraphics[width=0.48\textwidth]{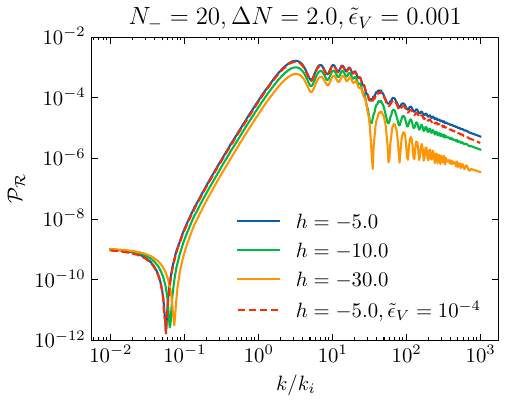}
    \caption{Power spectrum using the potential of
      eq.\eqref{eq:inf-pot-new}. Left: the duration $\Delta N$ of USR is
      varied. Right: the smoothness parameter $h$ is varied.}
    \label{fig:compare_DeltaN_h}
\end{figure}

After obtaining the background quantities from
eqs.~\eqref{eq:backgrounds}\footnote{One has to be careful while
  solving these equations: using a fixed size of the time step often
  leads to missing the start of USR phase by a small amount. Since the
  field is decelerating exponentially, this small difference could
  have visible impact on the power spectrum.}, the Mukhanov-Sasaki (MS)
equation~\eqref{eq:ms-eq} can be solved\footnote{We find the numerical
  treatment of ref.~\cite{karamPrimordialBlackHoles2023} useful.}, and
the power spectrum is computed using eq.~\eqref{eq:PS-u}. Some results
are shown in fig.~\ref{fig:compare_DeltaN_h}. The power spectrum has a
quite distinctive peak at a scale slightly larger than
$k_i = \left(aH \right)_i$, where the subscript $i$ denotes quantities
evaluated at the beginning of USR phase, i.e. at $\phi = \phi_i$. For
large wavelength ($k \ll k_i$), the power spectrum reduces to a near
scale-invariant one, as predicted by eq.~(\ref{eq:PS-SR}). At
$k \sim k_i/10$ the spectrum has a pronounced dip, followed by a power
law rise, $\mathcal{P}_\mathcal{R} \propto k^4$, which has been
observed e.g. in refs.~\cite{firouzjahiOneloopCorrectionsPower2023,
  piPrimordialBlackHole2023, byrnesSteepestGrowthPower2019}. The dip
can be explained by a cancellation between the two contributions to
$\mathcal{R}$ described by eq.~\eqref{eq:R-solution} while the second
term transitions from decaying to growing
\cite{piPrimordialBlackHole2023}.  After the peak, the power spectrum
shows oscillatory behavior, which can be traced back to the variation
of the power with time while the perturbation is still inside the
horizon; this is discussed in more detail in
appendix~\ref{app:power-evo}.

In the USR phase the inflaton field decelerates exponentially, and
fluctuations which exit the horizon around this phase grow
exponentially. Hence, the duration $\Delta N$ of the USR phase
strongly affects the maximal magnitude of the power spectrum at
$k \sim k_i$.  This is clearly shown in the left frame of
fig.~\ref{fig:compare_DeltaN_h}: increasing $\Delta N$ increases the
height of the peak. Since we fixed $N_- = 20$ (which implies
$N_+ = 35$), $\beta = 10^{-4}$ and $\tilde \epsilon_V = 10^{-3}$ in
this figure, varying $\Delta N$ requires variations of $\phi_e$.

Another important quantity for the power spectrum is the smoothness
parameter $h$. Intuitively, it determines how fast the inflaton field
goes back to the slow-roll, or $\epsilon_1 \to \epsilon_V$, after the
USR phase. The dependence on $h$ is explored in the right frame of
fig.~\ref{fig:compare_DeltaN_h}, where we fixed $\Delta N = 2$ for the
same choices of $\tilde\epsilon_V$ and $N_-$; different values of $h$
correspond to different $\beta$ (and also different $\phi_e$). We see
that $h$ largely determines the drop after the peak in the power
spectrum, with larger $|h|$ leading to a steeper drop. $h$ also has
some influence on the peak of the power spectrum, as has been shown in
ref.~\cite{firouzjahiInducedGravitationalWaves2023}; in particular,
$|h| \gg 1$ suppresses the peak somewhat. Through
eq.~\eqref{eq:smooth-h} the smoothness parameter $h$ depends on the
parameters $\Delta N$, $\tilde{\epsilon}_V$, and $\beta$. Therefore,
the different curves in the left frame also correspond to different
values of $h$, which affects the shape of the spectrum for
$k \gg k_i$.

On the other hand, we found that there almost is a degeneracy in
$\tilde{\epsilon}_V$ and $\beta$: the power spectrum remains
essentially unchanged if $\tilde{\epsilon}_V$ and $\beta$ are varied
so that $h$ is fixed. This can be seen by comparing the dashed red and
solid dark blue lines in the right frame, which have the same
smoothness parameter $h$: they indeed coincide except at very large
comoving momentum $k \gg k_i$. As we shall see, the same degeneracy
holds for the PBH abundance. 

In the subsequent sections, observables computed from the power
spectrum are compared with the corresponding bounds. We find a couple
of approximations can be applied to accelerate the computations. $N_-$
has no effects on the power spectrum \textit{around $k_i$}, other than
shifting the overall scale $k_i$. $\Delta N$ is the most important
parameter, but it doesn't change the shape of the spectrum around the
peak, as long as the smoothness parameter $h$ is fixed, see the left
frame in fig.~\ref{fig:compare_DeltaN_h}. To fair approximation we
can use interpolation to obtain the dependence of the power spectrum
on $N_-$ and $\Delta N$ using the results shown in
fig.~\ref{fig:compare_DeltaN_h}, without need of additional numerical
solution of the background and MS equations.

So far we have specified the dynamics of inflation and computed the
curvature perturbation power spectrum, which is defined via the
expectation value of the curvature perturbations, see
eq.~(\ref{eq:power_def}). However, the latter does not suffice to
specify the probability distribution of the fluctuations if there are
significant non-Gaussianities. We next turn to a computation of the
non-Gaussian curvature perturbations, which we need in order to
estimate the final PBH density.

%%%%%%%%%%%%%%%%%%%%%%%%%%%%%%%%%%%%%%%%%%%%%%%%%%%%%%%%%%%%%%%%%%%%%%%%%%%%%%%
\section{$\delta N$ calculation}
\label{sec:delta_N}

In this section we try to quantify the non-Gaussianity generated from
the USR phase of inflation. Local non-Gaussianity is one of the most
important shapes of non-Gaussianity. Phenomenologically, the following
parametrization for the curvature perturbation is often used to
specify such non-Gaussianity~\cite{baumannTASILecturesInflation2012}:
\begin{equation}    \label{eq:NG-def}
  \zeta(x) = \zeta_G(x) + \frac{3}{5} \fNL^{\text{local}} \left[
    \zeta_G(x)^2 - \expval{\zeta_G(x)^2} \right]\,.
\end{equation}
Here and hereafter we use the perturbation $\zeta$ rather than
${\cal R}$; as noted above, the two are identical for perturbations
(well) outside the Hubble horizon, which is true for all perturbations
of interest at the end of inflation. Moreover, $\zeta_G(x)$ denotes
the pure Gaussian curvature perturbation. In the subsequent discussion
we go beyond such quadratic local non-Gaussianity and derive the
relation between the full curvature perturbation and its Gaussian part
via the $\delta N$ formalism \cite{lythGeneralProofConservation2005,
  lythInflationaryPredictionPrimordial2005,
  sasakiGeneralAnalyticFormula1996}. To that end, we rewrite the
equation of motion in eq.~\eqref{eq:backgrounds} in terms of the
number of $e$-folds $N(t) = \ln a(t)$:
\begin{equation}    \label{eq:EOM-efold}
  \frac{\dd^2 \phi} {\dd N^2} + (3-\epsilon_1) \frac{\dd \phi} {\dd N}
  + \frac{V' (\phi)}{H^2(t)} = 0\,.
\end{equation}

\subsection{USR}
\label{sec:USR_only}

We first consider an (incomplete) scenario with only a single USR
phase of inflation described by $V(\phi) = V_0 = \const$.  The
equation of motion of the inflaton field, \eqref{eq:EOM-efold}, then
becomes
\begin{equation} \label{eq:eomusr}
\frac{\dd^2 \phi} {\dd N^2} + 3 \frac{\dd \phi} {\dd N} = 0\,.
\end{equation}
We use the potential shown in eq.~\eqref{eq:inf-pot-new}, where the
field starts at large values and the field velocity is always negative
during inflation. The field value $\phi$ and ``velocity''
$\pi \equiv \dd\phi / \dd N$ at the {\em end} of the USR phase
($t=t_e$) can be used as boundary conditions,
$(\phi_e, \pi_e) = (\phi(t_e), \pi(t_e))$. For convenience, we set
$N(t_e)=0$; so the number of $e$-folds prior to the end of USR would
be negative. The equation of motion (\ref{eq:eomusr}) has the solution
\begin{equation}	\label{math:USR_phi}
  \phi(N) - \phi_e = \frac{\pi_e}{3} \left( 1 - e^{-3N} \right)
  = \frac{1}{3} \left( \pi_e - \pi \right)\,.
\end{equation}
Thus, $N$ as function of field value and velocity can be obtained as
\begin{equation}    \label{eq:USR-N}
  N(\phi, \pi) = - \frac{1}{3} \ln( 1-\frac{\phi-\phi_e}{\pi_e/3})
  = -\frac{1}{3} \ln(\frac{\pi}{\pi_e})\,.
\end{equation}

Following the separate Universe approach, the curvature perturbation
can be calculated from the variation of the number of $e$-folds as
\cite{sasakiGeneralAnalyticFormula1996,
  lythGeneralProofConservation2005}\footnote{We choose the sign such
  that positive $\zeta$ corresponds to positive $\delta\phi$.}
\begin{equation}    \label{eq:Delta-N-formula}
\zeta = \delta N = N(\phi, \pi) - N(\phi+\delta\phi, \pi + \delta \pi)\,.
\end{equation}
It is important to note that $\phi_e$ is a constant, defined via the
inflaton potential (\ref{eq:inf-pot-new}). In contrast,
$\pi_e(\phi, \pi) = \pi + 3 (\phi - \phi_e)$ actually depends on $\phi$ and
$\pi$, c.f. eq.~\eqref{math:USR_phi}. The quantum fluctuation of the
field velocity $\delta \pi$ can be neglected
\cite{chenInin$dN$Calculations2013,
  caiHighlyNonGaussianTails2022}. For perturbations that exit the
horizon during USR at $\phi=\phi(t)$, the curvature perturbation can
be obtained from the last expression of eq.~\eqref{eq:USR-N}:
\begin{equation} \label{eq:zeta-pi}
\zeta = -\frac{1}{3} \ln(1 + \frac{3\delta\phi}{\bar\pi_e})\,.
\end{equation}
Here $\bar\pi_e$ is the velocity of the unperturbed background field, i.e. for
$\delta\phi = 0$. Recall that $\pi_e < 0$, so that $\delta\phi > 0$
implies $\zeta > 0$, as advertised.

Since $\phi$ is essentially a free quantum field, its perturbations
are Gaussian; however, since the relation between $\zeta$ and
$\delta\phi$ in eq.~(\ref{eq:zeta-pi}) is nonlinear, $\zeta$ has a
non-Gaussian distribution. One can also calculate the Gaussian part
of the curvature perturbation by considering the term linear in
$\delta \phi$ which is described by Gaussian statistics,
\begin{equation} \label{eq:zetag-pi}
  \zeta_G = \left| \pdv{N}{\phi}\right| \delta \phi
  = -\frac{\delta \phi}{\bar\pi_e}\,.
\end{equation}
Hence, the full non-linear curvature perturbation $\zeta$ can be written
in terms of its linear part $\zeta_G$ as
\cite{caiRevisitingNonGaussianityNonattractor2018,
  caiHighlyNonGaussianTails2022}
\begin{equation}	\label{math:trafo_only_USR}
\zeta = - \frac{1}{3} \ln(1 - 3 \zeta_G)\,.
\end{equation}
$\zeta_G$ follows a Gaussian distribution with standard deviation
$\Sigma_{YY}$, which will be defined in section \ref{sec:PBH}. By
conservation of probability, the probability distribution function
(PDF) of the full curvature perturbation is\footnote{Strictly speaking
  $P[\zeta]$ is properly normalized only if Gaussian fluctuations with
  $\zeta_G > 1/3$ do not contribute, since for such fluctuation
  $\zeta$ in eq.~(\ref{math:trafo_only_USR}) is ill-defined; or equivalently,
  if $\Sigma_{YY} \ll 1$. This is true for all parameters of interest; in
  fact, otherwise there might well be ``eternal inflation'', as
remarked towards the end of section~\ref{sec:setup}.}
\begin{equation}	\label{math:PDF_only_USR}
  P[\zeta] = P[\zeta_G(\zeta)] \left| \dv{\zeta_G}{\zeta} \right|
  = \frac{1}{\sqrt{2\pi \Sigma_{YY}}}
  \exp[- \frac{(1-\exp(-3\zeta))^2}{18\Sigma_{YY}} - 3 \zeta]\,.
\end{equation}
Hence, to first approximation $P[\zeta]$ reduces to a Gaussian
distribution function for $|\zeta| \lesssim \sqrt{\Sigma_{YY}} \ll 1$;
however, the probability for positive fluctuations with $\zeta \gtrsim
\sqrt{\Sigma_{YY}}$ is enhanced over that of a pure Gaussian.

\subsection{USR-SR transition}
\label{sec:usr-sr}

The potential has to have a conventional slow-roll (SR) phase after
the USR phase in order to avoid overproduction and to obtain the right
mass of PBHs. By also considering the transition, the relaxation to SR
can be tracked more carefully. Taking this phase into account will
modify the result of the previous subsection.

The potential in eq.~\eqref{eq:inf-pot-new} during the second SR phase
contains a cubic term; as explained above, this form of the potential
can be used for the entire duration of inflation, even for the
reheating epoch. Here we are only interested in the epoch immediately
following the USR phase. In order to have an analytical expression for
$\phi(N)$ to be used in the $\delta N$ computation, we expand the
potential around $\phi_e$ (with $\phi \leq \phi_e$):
\begin{equation} \label{eq:potexp}
  \frac{V(\phi)}{V_0} = 1 + \frac{\sqrt{2\epsilon_V(\phi_e)}}{\mpl}
  (\phi - \phi_e) + \frac{\eta_V(\phi_e)}{2\mpl^2}
  \left( {\phi-\phi_e} \right)^2 + \dots \,;
\end{equation}
$\epsilon_V(\phi_e)$ and $\eta_V(\phi_e)$ have been given in
eqs.~\eqref{eq:pot-SR-para-real}. We have checked that for the first
few $e$-folds of evolution after the USR phase, the predictions for
$\phi(N)$ derived from eq.~(\ref{eq:potexp}) closely match those
derived from the full potential eq.~\eqref{eq:inf-pot-new}.

Analogously to eq.~\eqref{eq:USR-N}, one can easily derive the
analytical solution $\phi(N)$ from eq.~(\ref{eq:potexp}). The total
number of $e$-folds from some point during the USR phase defined by
inflation field value $\phi$ and its velocity $\pi$ to some point deep
in the second slow-roll phase is found to be~\cite{caiRevisitingNonGaussianityNonattractor2018}
\begin{equation}     \label{eq:SR-phi}
  N(\phi, \pi) = \frac{-2}{s-3} \ln \left(\frac{\mpl}
    {-2\eta_V(\phi_e) \pi_e - 6 \mpl \sqrt{2\epsilon_V(\phi_e)}} \right)
  - \frac{1}{3} \ln(\frac{\pi}{\pi_e}) + \const.\,,
\end{equation}
where for convenience we introduce the constant 
$s \equiv \sqrt{9 - 12\eta_V(\phi_e)} \simeq 3 - 2
\eta_V(\phi_e)$. The additive constant in eq.~(\ref{eq:SR-phi}) has no
effect on the curvature perturbation in the $\delta N$ calculation.
After applying the $\delta N$ formula in
eq.~\eqref{eq:Delta-N-formula}, we obtain the curvature perturbation
\cite{caiRevisitingNonGaussianityNonattractor2018}
\begin{align} \label{math:USR_zeta_deltaphi}
  \zeta = \delta N
  &= N(\phi, \pi) - N(\phi + \delta\phi, \pi + \delta \pi)  \notag  \\
  &= \frac{-2}{s-3} \log \left(1+ \frac{1}{\bar\pi_e}
    \frac{6\eta_V(\phi_e) \delta \phi}{2 \eta_V(\phi_e) + h}\right)
    - \frac{1}{3} \log(1 + \frac{3 \delta\phi}{\bar\pi_e})\,.
\end{align}
The $\phi$-dependence of $N$ again solely comes from $\pi_e$, and the
smoothness parameter $h$ has been defined in
eq.~\eqref{eq:smooth-h}. We note that for $|h| \gg 1$ we recover the
USR-only scenario in eq.~\eqref{math:trafo_only_USR}. The reason is
that a large $|h|$, and hence large $|V'|$, speeds up the transition
to the slow-roll regime, where the dynamics is controlled by the first
derivative of the potential.

In comparison, the Gaussian part of the curvature perturbation is
given by
\begin{equation}	\label{eq:math:USR_SR_zeta_G}
  \zeta_G = \pdv{N}{\phi} \delta \phi = \frac{1}{\bar\pi_e}
  \left( \frac{-2\eta_V(\phi_e)}{s-3} \frac{6}{2\eta_V(\phi_e) + h} - 1 \right)
  \delta \phi =: \frac{3\alpha}{\bar\pi_e} \delta\phi\,.
\end{equation}
In the last step we have introduced the constant $\alpha$, defined by
\begin{equation}	\label{math:alpha-def}
  \alpha \equiv \frac{1}{3}\left( \frac{-2\eta_V(\phi_e)}{s-3}
    \frac{6}{2\eta_V(\phi_e) + h} - 1 \right)\,  .
\end{equation}
The full curvature perturbation can be expressed in terms of its
Gaussian part as
\begin{equation}	\label{eq:zeta_gaussian_relation}
  \zeta = \frac{-2}{s-3} \ln(1 + \frac{2\eta_V(\phi_e)}{2\eta_V(\phi_e) + h}
  \frac{\zeta_G}{\alpha}) - \frac{1}{3} \ln(1 + \frac{\zeta_G}{\alpha})\,.
\end{equation}

In the smooth transition limit $h \rightarrow 0$, one obtains
\begin{equation} \label{eq:alpha_smooth}
  \alpha = \frac{1}{3} \left( \frac{-6}{s-3} - 1 \right)
  \rightarrow \frac{1}{\eta_V(\phi_e)} - \frac{1}{3}\,,
\end{equation}
where last limit holds for $|\eta_V(\phi_e)| \ll 1$. Moreover, for
$h \rightarrow 0$ the relation between the Gaussian and full curvature
perturbations simplifies to
\begin{equation}	\label{math:trafo_smooth}
\zeta = \alpha \ln(1 + \frac{\zeta_G}{\alpha})\,.
\end{equation}
In this limit one can also easily find an analytical expression for
$\zeta_G(\zeta)$, so that the probability distribution $P[\zeta]$ can
be computed analytically. On the other hand,
eq.~(\ref{eq:zeta_gaussian_relation}) cannot be solved analytically for
$\zeta_G$. Fortunately, we will see below that the resulting PBH density
can be computed by directly integrating over $\zeta_G$.

Before concluding this section, we note that expanding
eq.~\eqref{eq:zeta_gaussian_relation} around $\zeta_G = 0$ yields the
non-Gaussianity parameter defined in eq.~\eqref{eq:NG-def}
\cite{caiRevisitingNonGaussianityNonattractor2018}:
\begin{equation}	\label{eq:f_USR}
	\fNL = \frac{5h (h-\eta_V(\phi_e))}{2(h-6)^2}.
\end{equation}
This becomes maximal for $|h| \rightarrow \infty$. In case of a very
rapid transition from the USR to the second slow-roll phase an order
of unity quadratic non-Gaussian parameter can thus be achieved,
$\fNL \rightarrow 5/2$.

In this section, we have derived the relation between the full
curvature perturbation and its Gaussian part using the $\delta N$
formalism. We will now show how such non-Gaussianity affects the
PBH abundance.

\section{Observables} \label{sec:PBH}

In previous sections, the formalism for the computation of the power
spectrum and non-Gaussian curvature perturbations have been
presented. In this section, we aim to determine the resulting PBH
abundance as well as the spectrum of gravitational waves associated
with PBH formation.

\subsection{PBH mass function and abundance}
\label{sec:pbhm}

As usual, we assume spherical collapse when estimating the PBH
abundance. This depends on the curvature perturbation $\zeta(\vec x)$
integrated over a sphere. When applying a multipole expansion to
$\zeta(\vec x)$ this integration projects out the monopole
$\zeta_0(r)$ which depends only on the distance $r$ from the center of
the sphere; the higher multipoles do not contribute. In Fourier space
this corresponds to $\zeta_0(k)$, which depends only on the absolute
value of the wave vector $k = \left|\vec k \right|$. Recall that
the power spectrum also only depends on $k$, see eq.~(\ref{eq:PS-u}).

It has been argued that the compaction function $\mathcal{C}$ is the
most crucial quantity for determining the PBH formation
\cite{escrivaPrimordialBlackHoles2022,
  gowNonperturbativeNonGaussianityPrimordial2022,
  delucaNoteAbundancePrimordial2022,
  muscoThresholdPrimordialBlack2019,
  ferrantePrimordialNongaussianityAll2023}. It is essentially the
volume-averaged density contrast $\delta$
\cite{muscoThresholdPrimordialBlack2019}. For small curvature
perturbation $\zeta$, a linear relation between curvature perturbation
and density contrast in Fourier space can be derived
\cite{muscoThresholdPrimordialBlack2019} (see also appendix
\ref{sec:app_var}):
\begin{equation}  \label{eq:delta-zeta-linear}
  \delta(k) \approx - \left( \frac{k}{aH} \right)^2
  \frac{2(1+\omega)}{5+3\omega} \zeta_0(k).
\end{equation}
This is not sufficient when a highly non-linear process such as PBH
formation is considered~\cite{muscoThresholdPrimordialBlack2019,
  delucaNoteAbundancePrimordial2022}. A more reliable method to
compute the PBH abundance was proposed in
refs.~\cite{gowNonperturbativeNonGaussianityPrimordial2022,
  ferrantePrimordialNongaussianityAll2023,
  delucaNoteAbundancePrimordial2022}. It is based on the full relation
between the compaction function $\mathcal{C}$ and the curvature
perturbation $\zeta$ \cite{muscoThresholdPrimordialBlack2019,
  gowNonperturbativeNonGaussianityPrimordial2022}:
\begin{equation} \label{eq:Cofr}
  \mathcal{C}(r) = - \frac{2}{3} r \zeta_0'(r) \left[ 2 + r \zeta_0'(r)
  \right]\,,
\end{equation}
where $\zeta'_0 = \dd\zeta_0 / \dd r$.

The first step is to calculate the probability distribution of the
linear compaction function $\mathcal{C}_l$, defined as
\begin{equation}    \label{eq:linear_compaction}
\mathcal{C}_l \equiv -\frac{4}{3} r \zeta_0'\,.
\end{equation}
We saw in the previous section that $\zeta$ in general depends
nonlinearly on the Gaussian perturbation $\zeta_G$. As a result,
$\mathcal{C}_l$ depends on two random variables: $X := r \zeta'_{0, \,G}$
and $Y:=\zeta_{0, \, G}$.
Ref.~\cite{gowNonperturbativeNonGaussianityPrimordial2022} showed that
the PDF for $\mathcal{C}_l$ can be calculated as a marginal
probability:
\begin{align}	\label{math:PDF_C_l}
  P[\mathcal{C}_l]
  &= \int \dd{\zeta_{0, \, G}} \frac{3}{4 |\mathcal{J}_1(\zeta_{0 ,\, G})|}
    P \left[ -\frac{1}{\mathcal{J}_1(\zeta_{0,\,G})}
    \frac{3}{4} \mathcal{C}_l\,  ,\, \zeta_{0,\,G} \right]\, .
\end{align}
We have introduced the auxiliary quantity (Jacobian)
\begin{equation}	\label{eq:J}
  \mathcal{J}_1(\zeta_{0,\,G}) = \pdv{\zeta_0}{\zeta_{0,\,G}}
  = \frac{P_G[\zeta_{0,\,G}]}{P[\zeta_0(\zeta_{0,\,G})] }\,,
\end{equation}
and used the fact that in our calculation the full curvature
perturbation $\zeta$ depends on $\vec x$ only through the
$\vec x$-dependence of $\zeta_G$, see eqs.~(\ref{math:trafo_only_USR})
and (\ref{eq:zeta_gaussian_relation}). Moreover, $P(X,Y)$ in
eq.~(\ref{math:PDF_C_l}) is a 2D Gaussian distribution; the entries of
its $2\times 2$ covariance matrix are given by:
\begin{align} \label{eq:variances}
    \begin{split}
      \Sigma_{XX} &= 	\int \dd{(\ln k)} (kr)^2 j_1^2(kr)
      \mathcal{P}_{\mathcal{R}}(k)\,,    \\
      \Sigma_{XY} &= 	\int \dd{(\ln k)} (kr) j_0(kr) j_1(kr)
      \mathcal{P}_{\mathcal{R}}(k)   \,, \\
     \Sigma_{YY} &= \int \dd{(\ln k)} j_0^2 (kr) \mathcal{P}_{\mathcal{R}}(k)\,.
    \end{split}
\end{align}
Here $j_0(x) = \sin(x)/x$ and $j_1(x) = \sin(x)/x^2 - \cos(x)/x$ are
spherical Bessel functions, and the power spectrum has been given in
eq.~(\ref{eq:PS-u}); see appendix~\ref{sec:app_var} for further
details.

It is often mentioned that the non-Gaussian tail of the PDF $P[\zeta]$
enhances PBH production. Despite the complexity of
eq.~\eqref{math:PDF_C_l}, one can still qualitatively understand the
relation between the tails of $P[\zeta]$ and $P[\mathcal{C}_l]$. The
non-Gaussian statistics appears in $P[\mathcal{C}]$ only via the
Jacobean $\mathcal{J}_1$, which can be computed from the function
$\zeta(\zeta_G)$, see eqs.~\eqref{math:trafo_only_USR} or
\eqref{eq:zeta_gaussian_relation}. The factor $|\mathcal{J}_1|^{-1}$
outside the 2D Gaussian distribution in eq.~\eqref{math:PDF_C_l} is
usually not very important, but the same factor inside the argument
can change the PDF considerably.

For illustration let us consider the much simpler case where only the
direct USR contribution is considered, see section~\ref{sec:USR_only}.
From eq.~(\ref{math:trafo_only_USR}) we have
$\mathcal{J}_1 = ( 1 - 3 \zeta_{0,\,G})^{-1}$, hence the 2D Gaussian
distribution in eq.~\eqref{math:PDF_C_l} is
$\sim {P[-\frac{3}{4}\mathcal{C}_l (1-3\zeta_{0, \, G}), \zeta_{0, \,G}]}$. It
still has maximal probability when both arguments vanish, but the
factor $\mathcal{J}_1^{-1}$ enhances the distribution (more exactly,
leads to less suppression) for positive fluctuations $\zeta_{0, \, G}$. Not
surprisingly, $\mathcal{J}_1^{-1}$ is also linked to the non-Gaussian
tail found in $P[\zeta]$ for this case, see
eq.~\eqref{math:PDF_only_USR}.

\begin{figure}[ht]
\centering
\includegraphics[width=0.95\textwidth]{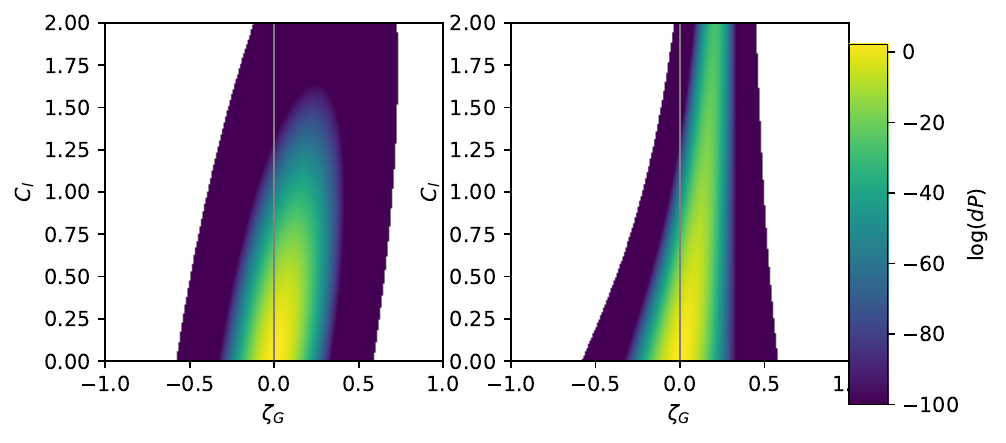}
\caption{The integrand of eq.~\eqref{math:PDF_C_l}. The left frame
  shows the Gaussian case ($\mathcal{J}_1 = 1$), while the right frame
  shows the pure USR case ($\mathcal{J}_1 = (1-3\zeta_{0,\,G})^{-1}$).}
\label{fig:P_C_l_integrand}
\end{figure}

In fig.~\ref{fig:P_C_l_integrand} the integrand of
eq.~\eqref{math:PDF_C_l} is shown as a function of $\mathcal{C}_l$ and
$\zeta_{0. \, G}$; the PDF $P[\mathcal{C}_l]$ can then be obtained by
integrating along the $\zeta_{0, \,G}$ direction. The pure Gaussian case,
$\mathcal{J}_1 = 1$, is shown on the left, and on the right the pure
USR case with $\mathcal{J}_1 = (1-3\zeta_{0, \, G})^{-1}$. We see that the
probability is enhanced at sizable, positive
$\zeta_{0, \, G} < 1/3$\footnote{Note that both the first argument of the
  Gaussian and the factor in front of the Gaussian vanish if
  $\mathcal{J}_1^{-1} = 0$, i.e. for $\zeta_{0, \, G} = 1/3$.}; the
enhancement is most prominent for larger $\mathcal{C}_l$. The
``spread'' in $\mathcal{C}_l$-direction also depends on the variance
$\Sigma_{XX}$, with larger $\Sigma_{XX}$ leading to sizable
probability for larger $\mathcal{C}_l$; we will now discuss how this
enhances the PBH abundance.

The initial PBH mass can be related to the horizon mass $M_H$ at the
time when a sufficiently large curvature perturbation re-entered the
horizon \cite{carrPrimordialBlackHoles2020,
  escrivaPrimordialBlackHoles2022}:
\begin{equation} \label{eq:mPBH}
m_{\text{PBH}} = \kappa M_{H} \equiv \kappa \frac{4\pi\mpl^2}{H}\, .
\end{equation}
As already noted, we are interested in PBHs in the asteroid mass
range, which could form all of dark matter. We also already mentioned
that our model of inflation requires trans-Planckian field values.
The Hubble parameter during inflation is therefore quite large,
$H \sim 0.4 \cdot 10^{-3}\sqrt{\tilde \epsilon_V} \; \mpl$. This
usually leads to a rather large reheat temperature, in which case
perturbations giving rise to asteroid mass PBHs reentered during the
radiation dominated epoch, where the equation of state parameter
$\omega = 1/3$, and the numerical factor $\kappa$ appearing in
eq.~(\ref{eq:mPBH}) is found by a simple analytical method to be
$\sim 0.2$~\cite{carrPrimordialBlackHole1975}. Depending on the shape
of the fluctuation, the PBH mass can grow due to accretion,
conceivably leading to values larger than $M_{H}$
\cite{bicknellFormationPrimordialBlack1979,
  carrNewCosmologicalConstraints2010,
  escrivaSimulationPrimordialBlack2020}. On the other hand, according
to ref.~\cite{choptuikUniversalityScalingGravitational1993} critical
phenomena limit this growth, leading to the following mass {\em after}
the completion of accretion:
\begin{equation}    \label{eq:BH_M_scaling}
  m = K M_{H} (\mathcal{C} - \mathcal{C}_c)^\gamma
  = K M_H \left( \mathcal{C}_l - \frac{3}{8}\mathcal{C}_l^2
    - \mathcal{C}_c \right)^\gamma\,.
\end{equation}
Here the exponent $\gamma$ depends on the equation of state parameter
$\omega$, and the $\order{1}$ factor $K$ weakly depends on the shape
of the curvature fluctuation that triggers the collapse as well as on
$\omega$ \cite{escrivaPrimordialBlackHoles2022}. Evidently only large
perturbations, corresponding to a compaction function larger than the
critical value $\mathcal{C}_c$, lead to PBH formation. Following
ref.~\cite{gowNonperturbativeNonGaussianityPrimordial2022}, $K = 1$,
$\gamma=0.36$ and $\mathcal{C}_c = 0.587$ are used. Note that
$\mathcal{C} = \mathcal{C}_l - 3 \mathcal{C}_l^2/8$ reaches a maximal
value of $\mathcal{C}_{\rm max} = 2/3$ at $\mathcal{C}_l = 4/3$;
according to eq.(\ref{eq:BH_M_scaling}) this corresponds to a
maximal PBH mass well below the horizon mass,
\begin{equation} \label{eq:mmax}
  m_{\rm max} = 0.4 M_H\,.
\end{equation}
Moreover, $\mathcal{C} > 0.587$ for $\mathcal{C}_l \in [0.872,1.794]$.

In our numerical estimates we evaluate $\mathcal{C}$ and its PDF at
$r = 2.74/k_m$, where $k_m$ is the location of the absolute maximum of
the power spectrum $\mathcal{P}_{\mathcal{R}}(k)$; this choice
maximizes the PBH abundance for a monochromatic power spectrum
\cite{muscoThresholdPrimordialBlack2021,
  gowNonperturbativeNonGaussianityPrimordial2022}.

The Press-Schechter ansatz states that a perturbation exceeding the
threshold value $\mathcal{C}_c$ will collapse into a black hole after
horizon re-entry~\cite{pressFormationGalaxiesClusters1974}. The
scaling law~\eqref{eq:BH_M_scaling} determines the final PBH mass from
the compaction function; the PBH mass and abundance together determine
the final PBH mass density~\cite{escrivaPrimordialBlackHoles2022}. The
PBH abundance can be given by integrating the probability distribution~\cite{escrivaPrimordialBlackHoles2022}
\begin{equation}     \label{eq:PS-formula}
   \frac{\rho_\text{PBH}}{\rho_{\rm tot}}
  = 2 \int_{\mathcal{C}_c}^{\mathcal{C}_{\rm max}} \dd{\mathcal{C}} K
  (\mathcal{C} - \mathcal{C}_c)^{\gamma} P[\mathcal{C}]\,.
\end{equation}
Here the PDF
\begin{equation} \label{eq:PC}
P[\mathcal{C}] = P[\mathcal{C}_l] \left( \dd \mathcal{C} / \dd
  \mathcal{C}_l \right)^{-1}
= P[\mathcal{C}_l] \cdot \frac {1} {1 - 3 \mathcal{C}_l /4}\,,
\end{equation}
where $P[\mathcal{C}_l]$ has been given in eq.~(\ref{math:PDF_C_l}). The
PBH mass function is defined as the fractional contribution of PBHs to
the total scaled cold dark matter density per logarithmic mass
interval~\cite{carrPrimordialBlackHole2017}:
\begin{equation}    \label{eq:extended_f}
    f(m) \equiv \frac{1}{\Omega_{\text{CDM}}} \dv{\Omega_\text{PBH}}{\ln m}\,.
\end{equation}
Like all matter densities, the energy density of PBHs formed in the
radiation epoch grows linearly with the scale factor $a(t)$ relative
to the radiation energy density until matter radiation equality is
reached; we express this factor using the ratio of horizon masses
$M_H/M_{H, \text{eq}}$. Hence, the PBH mass function can be written as
\cite{gowNonperturbativeNonGaussianityPrimordial2022,
  ferrantePrimordialNongaussianityAll2023}
\begin{align}\label{eq:f_m_full}
 f(m) &= \frac{2}{\Omega_\text{CDM}} K (\mathcal{C} - \mathcal{C}_c)^\gamma
         \dv{\mathcal{C}}{\ln m} \lrfrac{M_H}{M_{H, \, \eq}}^{-1/2}
         P[\mathcal{C}] \notag \\
      &= \frac{2}{\Omega_\text{CDM}} \lrfrac{M_H}{M_{H, \, \eq}}^{-1/2} K
        \frac{\left(\mathcal{C}_l - \frac{3}{8} \mathcal{C}_l^2
        - \mathcal{C}_c \right)^{\gamma+1}}{\gamma
        \left(1-\frac{3}{4}\mathcal{C}_l \right)} P[\mathcal{C}_l]\,,
\end{align}
where again $\mathcal{C} > \mathcal{C}_c$ is required and
eq.~(\ref{eq:PC}) has been used. The total PBH abundance can be
computed through a simple integration
\begin{equation} \label{eq:fPBH}
f_{\text{PBH}} = \int_0^{m_\text{max}} \dd{(\ln m)} f(m)\,,
\end{equation}
where the maximal PBH mass $m_{\rm max}$ has been given in
eq.~(\ref{eq:mmax}).\footnote{In this discussion we restricted
  ourselves to so-called type-I perturbation. Type-II perturbations
  are not well studied, but are most likely suppressed
  \cite{escrivaPrimordialBlackHoles2022}.}

The horizon mass $M_H$ corresponding to a mode that exits the horizon
at the end of USR can be computed from the parameters the model of
inflation. At the horizon exit (during inflation), a certain mode can
be expressed in terms of the Hubble scale at the end of inflation
$H_\text{end}$ and the duration of the second SR phase:
$k = aH = (k/k_e) \, a_\text{end} H_\text{end} \, e^{-N_-}$, where the
subscript ``end'' denotes the end of inflation, not the end of
USR. Assuming instantaneous reheating, the Universe enters radiation
domination immediately so that $H \propto a^{-2}$ thereafter. Then,
the Hubble parameter at the time when this mode re-enters the horizon
is $H_\text{re-enter} = H_\text{end} (k/k_e)^2 e^{-2N_-}$. Thus, the
horizon mass can be related to the number of $e$-folds of the second
SR phase $N_-$~\cite{garcia-bellidoGravitationalWavesInterferometer2016}:
\begin{equation}    \label{eq:horizon_mass_N}
  M_H (k) = \frac {4\pi\mpl^2} {H_\text{end}} \lrfrac{k}{k_e}^{-2} e^{2N_-}
  \simeq \num{2.51e-38} \frac{\mpl}{H_\text{end}} \lrfrac{k}{k_e}^{-2}
  e^{2N_-} M_\odot\,,
\end{equation}
where we have neglected the variation of the Hubble parameter during
inflation. The horizon mass entering eq.~\eqref{eq:f_m_full} can be
evaluated at the scale $k_m \gtrsim k_i$ where the power spectrum
reaches its maximum, see fig.~\ref{fig:compare_DeltaN_h}. For an
asteroid mass PBH,
$M_{\text{PBH}} \simeq M_H \simeq 10^{-15} M_\odot$, the duration of
the second SR phase amounts to $N_- \simeq 24 + \log(H_e / \mpl)/2$.

\subsection{Scalar Induced Gravitational Waves}

We saw that PBH formation requires large density perturbations;
gravitational waves are then inevitably generated at second
order. They basically do not interact after production,
so they can allow us to gain knowledge of the epoch when PBHs were
formed.

The FLRW metric together with the scalar perturbation $\Phi$ as well
as the tensor perturbation $h_{ij}$ in conformal Newtonian gauge in
conformal time can be written as
\cite{kohriSemianalyticCalculationGravitational2018,
  gorbunovIntroductionTheoryEarly2011}
\begin{equation} \label{eq:pert-metric}
  \dd{s}^2 = a^2(\eta) (1+2\Phi) \dd{\eta}^2 - a^2(\eta) \left(
    (1-2\Phi)\delta_{ij} + \frac{1}{2}h_{ij} \right)\dd{x^i} \dd{x^j}\,.
\end{equation}
A possible anisotropic stress has been neglected. Perturbations with
different helicities are decoupled only at first order. At second
order, tensor perturbations can be generated from scalar perturbations
\cite{kohriSemianalyticCalculationGravitational2018}, schematically:
\begin{equation} \label{eq:gw1}
  \mathcal{P}_h = \int \dd{k} \int \dd{k'} \left( \int \dd{t} f(k, k', t)
  \right)^2 \mathcal{P}_\mathcal{R}(k) \mathcal{P}_\mathcal{R}(k')\,,
\end{equation}
where $f(k, k', t)$ is some oscillating function. For PBHs of asteroid
mass, the perturbations re-enter the horizon during radiation
domination and the so-called scalar induced gravitational waves (SIGW)
are produced around the re-entry as well. The differential energy
density of these gravitational waves today,
$\Omega_{\rm GW,0}(k) := \dd \Omega_{\rm GW,0} / \dd \log(k)$, is given by
\cite{espinosaCosmologicalSignatureSM2018,
  witkowskiSIGWfastPythonPackage2022,
  anandaCosmologicalGravitationalWave2007,
  baumannGravitationalWaveSpectrum2007,
  kohriSemianalyticCalculationGravitational2018,
  escrivaPrimordialBlackHoles2022,
  domenechScalarInducedGravitational2021}
\begin{equation}    \label{eq:SIGW-spec}
  \Omega_\text{GW,0} (k) = c_g \Omega_{r, 0} \int_0^1 \dd{d}
  \int_1^{\infty} \dd{s} \mathcal{T}_{\text{rad}}(d,s) \mathcal{P}_\mathcal{R}
  \left( \frac{k}{2} (s+d) \right)
  \mathcal{P}_\mathcal{R} \left( \frac{k}{2} (s-d) \right)\,,
\end{equation}
with kernel function
\begin{equation} \label{eq:kernel}
  \mathcal{T}_{\text{rad}}(d,s) = 12 \frac {(d^2 -1)^2 (s^2 - 1)^2
    (d^2 + s^2 - 6)^4} {(s^2 - d^2)^8} \left[
    \left( \ln \frac {3-d^2} {|s^2 -3|} + \frac {2(s^2 -d^2)}{d^2 + s^2 -6}
    \right)^2 + \pi^2 \theta(s - \sqrt{3}) \right]\,.
\end{equation}
Here the constant $c_g$ relates the effective number of relativistic
degrees of freedom during GW production to that today:
\begin{equation*}
  c_g \equiv \frac {g_{*, \text{rad}}} {g_{*, 0}} \left(
    \frac{g_{*s, 0}}{g_{* s, \text{rad}}} \right)^{4/3}\,.
\end{equation*}
If perturbations collapsing into asteroid mass PBHs reentered the
horizon in the radiation dominated universe, as we have assumed, the
corresponding temperature was so high that all degrees of freedom in
the Standard Model were relativistic; this leads to $c_g \approx 0.4$.
Finally, $\Omega_{r, 0} = \num{1.62e-5}$ in eq.(\ref{eq:SIGW-spec}) is
the current radiation energy density fraction \cite{planck6}.

Equation~\eqref{eq:SIGW-spec} only accounts for GW production at the lowest
nontrivial order with the curvature perturbation power spectrum as the
input. Primordial non-Gaussianity will not alter SIGW significantly,
unless $f_{\text{NL}} \gg 1$
\cite{adsheadNonGaussianityInducedGravitational2021,
  liPrimordialNonGaussianityF_2023,
  inuiConstraintsNonGaussianPrimordial2024}. As we have maximally
$\fNL = 5/2$, see eq.~(\ref{eq:f_USR}), this effect can safely be
neglected in our scenario.

Analogously to eq.~\eqref{eq:horizon_mass_N}, the present frequency of
the SIGW can be expressed in terms of $N_-$ as
\begin{equation}  \label{eq:f_N_-}
  f(k) = \frac{k}{2\pi a_0} = \num{3.62e10} \lrfrac{H_{\text{end}}}{\mpl}^{1/2}
  \frac{k}{k_i} e^{-N_-} \; \si{\hertz},
\end{equation}
where we have taken the effective degrees of freedom at GW production
to be $g_* = 106.75$, while today $g_{*, 0} = 3.91$, and the present
photon (CMB) temperature $T_0 = \SI{2.35e-4}{\eV}$ is used as well ~\cite{planck6}. 
For PBHs of mass $\sim 10^{-15}M_\odot$ the associated
GWs peak at frequency $f = \order{0.1} \si{\hertz}$.

In this section, we have established the formalism to compute the PBH
mass and abundance and the associated production of GWs. Next, numerical
results for our model will be presented.

\section{Results}
\label{sec:result}

\begin{figure}[ht]
\centering
\begin{subfigure}[t]{0.49\textwidth}
 \centering
  \includegraphics[width=\textwidth]{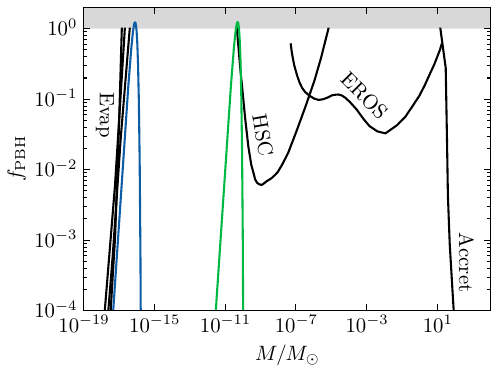}
  \caption{PBH bounds (black) from
    ref.~\cite{carrPrimordialBlackHoles2026} and potential PBH spectra
    predicted by our model (in color).}
\label{fig:f_bounds_h=-20}
\end{subfigure}
\hfill
\begin{subfigure}[t]{0.49\textwidth}
 \centering
  \includegraphics[width=\textwidth]{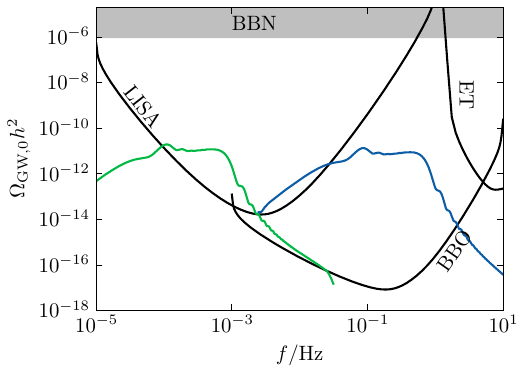}
  \caption{Future GW sensitivity curves (black) from
    ref.~\cite{schmitzNewSensitivityCurves2021} and potential SIGW
    signals predicted by our model (in color).}
\label{fig:GW_Omega-h=-20}
\end{subfigure}
\caption{PBH mass spectrum and the associated scalar induced GW
  signals with smoothness parameter $h=-20$ and
  $\tilde{\epsilon}_V = 10^{-4}$. The blue (green) curves correspond
  to the lightest (heaviest) PBH spectrum allowed in our model if PBHs
  form all DM, hence to the smallest (largest) allowed values of
  $N_-$.}
\label{fig:all_h=-20}
\end{figure}

Because of the critical scaling in eq.~\eqref{eq:BH_M_scaling}, the
PBHs do not have the exact same masses. Constraints on the PBH
abundance are usually formulated for monochromatic mass functions,
instead of an extended mass function defined in
eq.~\eqref{eq:extended_f}. An extended mass function is consistent with
the existing constraints on the PBH abundance if
\cite{carrPrimordialBlackHole2017, bellomoPrimordialBlackHoles2018}
\begin{equation}    \label{eq:check_BH_mass_f}
  \int \dd{\ln(m)} \frac{f(m)}{f_\text{max}(m)} \leq 1\,,
\end{equation}
where $f_\text{max}(m)$ is the maximally allowed monochromatic PBH
fraction at mass $m$. There are numerous constraints with varying
degrees of validity. In this work we choose the most well-established
bounds on the asteroid mass PBHs, which come from upper bounds on the
Hawking evaporation and on microlensing. The constraints are taken
from ref.~\cite{carrPrimordialBlackHoles2026}. The microlensing
constraints come mainly from Hyper-Supreme Cam (Subaru/HSC)
\cite{smythUpdatedConstraintsAsteroidMass2020} and EROS
\cite{tisserandLimitsMachoContent2007}. PBH evaporation contributes to
the background fluxes of energetic photons
\cite{carrNewCosmologicalConstraints2010} as well as positrons
\cite{boudaudVoyager1$e^pm$2019}. PBHs evaporating around
recombination could also alter CMB anisotropies
\cite{acharyaCMBBBNConstraints2020}. These evaporation constraints are
very close to each other and define the lower end of the allowed PBH
window shown in fig.~\ref{fig:f_bounds_h=-20}. The
bound on modifications of the ionization history due to DM halo
accretion, derived from CMB observations
\cite{serpicoCosmicMicrowaveBackground2020}, is included as well; this
is shown on the right edge of fig.~\ref{fig:f_bounds_h=-20} as
``Accret''. If one wishes PBH to form all of dark matter in the
Universe with relatively narrow mass spectrum, this window around
$\sim 10^{-14} M_\odot$ is the only option. As already noted, this
requires around $20$ $e$-folds of inflation after the USR phase.

From section~\ref{sec:PS}, we see that $\Delta N$ mostly affects the
normalization of the power spectrum around the peak, while the shape
of the spectrum at $k \gtrsim k_i$ is controlled by the smoothness
parameter $h$. Hence, for any given $h$, the duration $\Delta N$ can
be adjusted so that $f_\text{PBH} = 1$ is ensured. We can then vary
the duration of the second SR phase $N_-$ to achieve the desired PBH
mass density. The heaviest and lightest scenarios that are still
compatible with the current bounds via eq.~\eqref{eq:check_BH_mass_f},
with $\tilde{\epsilon}_V = 10^{-4}$ and $h=-20$, are shown as colored
curves in fig.~\ref{fig:all_h=-20}. Naturally, all values of $N_-$
in between these two extreme cases are also compatible with PBHs
forming all DM.

This PBH mass spectrum can be correlated with the SIGW. This is shown
in fig.~\ref{fig:GW_Omega-h=-20}, along with the projected sensitivity
curves of Laser Interferometer Space Antenna (LISA), Big Band
Observatory (BBO) and Einstein Telescope (ET), taken from
ref.~\cite{schmitzNewSensitivityCurves2021}. The BBN bound on
additional light degrees of freedom $\Delta N_\text{eff}$ (which
includes GWs, of course) is shown as the gray band on top of the
figure: $\Omega_{\text{GW}, 0} h^2 \lesssim 10^{-6}$
\cite{capriniCosmologicalBackgroundsGravitational2018}. Note that
heavier PBHs correspond to lower GW frequencies
($k \propto f \propto M^{-1/2}$): the green curve in the left frame of
fig.~\ref{fig:all_h=-20} is associated with the green curve in the
right frame.

The results presented so far indicate that these GW signals should be
easily detectable by LISA and/or BBO. However, the sensitivity curves
shown are projected using GW signals expected from first-order phase
transitions \cite{schmitzNewSensitivityCurves2021}. Direct comparison
between the signals (colored curved) and such sensitivity curves
(black curves) is not well justified, since the two GW signals have
different spectra. A safer criterion of detectability is based on the
signal-to-noise (SNR) ratio \cite{schmitzNewSensitivityCurves2021,
  papanikolaouGravitationalWavesUniverse2021}
\begin{equation}    \label{eq:SNR}
  \text{SNR}^2 = n_\text{det} t_\text{obs} \int \dd{f}
  \lrfrac{\Omega_\text{signal}(f)}{\Omega_\text{noise}(f)}^2\,,
\end{equation}
where $n_\text{det} = 1 (2)$ for an auto-correlation (cross-correlation)
type detector and $t_\text{obs}$ specifies the duration of the observation,
which is conservatively taken to be $1\; \si{\year}$ here.

In section~\ref{sec:usr-sr} we saw that the non-Gaussianity increases
with increasing $|h|$, i.e. for more rapid transition from the USR
phase to the second slow-roll epoch. Moreover, in section~\ref{sec:pbhm}
it was shown that larger non-Gaussianity increases the likelihood of
large density fluctuations for fixed power. As a result, larger $|h|$
requires less curvature perturbation on the power spectrum
level\footnote{We saw in the right frame of
  fig.~\ref{fig:compare_DeltaN_h} that increasing $|h|$ while keeping
  all other parameters of the potential fixed reduces the power
  spectrum near $k_m$. This leads to a {\em decrease} of
  $f_{\rm PBH}$, which has to be compensated by a (small) increase of
  $\Delta N$ if $f_{\rm PBH}$ is to be kept fixed. However, the
  resulting power near the maximum is still lower than for small
  $|h|$. If instead the power at $k_m$ was held fixed while $|h|$ is
  increased, we find that the PBH abundance would increase by about
  ten orders of magnitude when $|h|$ is increased from $2$ to $20$.}
in order to achieve $f_\text{PBH}=1$. On the other hand, the SIGW
signal receives contributions from a much broader range of curvature
perturbations; there is no strong threshold effect. The probability of
these smaller fluctuations is less enhanced by the
non-Gaussianity. The upshot of this discussion is that the associated
GW signals are {\em reduced} by non-Gaussianity if $f_\text{PBH}$ is
kept fixed. Moreover, $h$ also affects the shape of the power spectrum
after the peak, with larger $|h|$ leading to a steeper decrease at
large $k$, see the right frame of
fig.~\ref{fig:compare_DeltaN_h}. Larger $|h|$ therefore also leads to
a steeper slope of the GW signal at high frequency.

\begin{figure}[ht]
    \centering
    \begin{subfigure}[t]{0.48\textwidth}
    \begin{center}
    \includegraphics[width=\textwidth]{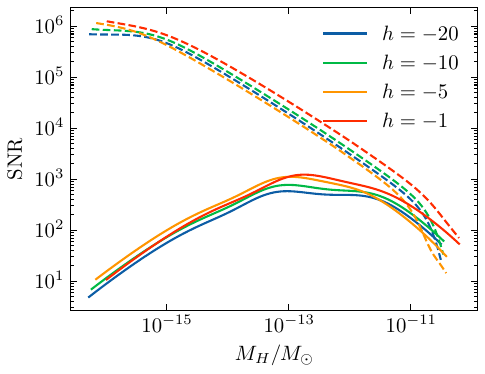}
    \end{center}
    \caption{With $\tilde{\epsilon}_V=10^{-4}$.}
    \label{fig:SNR_N2_1}
    \end{subfigure}% 
    \begin{subfigure}[t]{0.48\textwidth}
    \begin{center}
    \includegraphics[width=\textwidth]{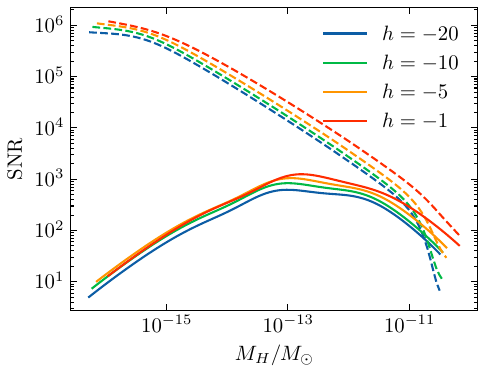}
    \end{center}
    \caption{With $\tilde{\epsilon}_V = 10^{-3}$.}
    \label{fig:SNR_N2_2}
    \end{subfigure}
    \caption{Signal-to-noise ratio at LISA (solid lines) and BBO
      (dashed lines) against the horizon mass upon re-entry of the
      mode $k_m$, for four values of the smoothness parameter
      $h$. $\Delta N$ is adjusted such that $f_{\rm PBH} = 1$.}
    \label{fig:SNR_N2}
\end{figure}

Both effects contribute to the SNR curves shown in
fig.~\ref{fig:SNR_N2}, where the projected SNRs at LISA (solid lines)
and BBO (dashed lines), computed using eq.~\eqref{eq:SNR}, are plotted
against the horizon mass at horizon crossing $M_H$, related to $N_-$
via eq.~\eqref{eq:horizon_mass_N}. The range of $M_H$ is chosen so that
the produced PBHs fit into the asteroid window: the range depends
slightly on $h$ as the PBH mass functions have different shapes by
eq.~\eqref{eq:f_m_full}. We see that all scenarios have SNR at both
experiments (much) larger than unity. Increasing $|h|$ generically
reduces the SNR, due to the NG effects enhancing PBH formation.

There are also changes in the shapes of the SNR curves when $h$ is
varied; this is due to the change of the high frequency tail. For
example, for $M_H \sim 10^{-13}M_\odot$ the curves showing the SNR at
LISA flatten, as the GW signals are all contained in the detection
range anyway. However, we saw in fig.~\ref{fig:GW_Omega-h=-20} that
for larger PBH mass, LISA is more sensitive to the higher end of the
GW spectrum, which gets suppressed when $|h|$ is increased. As a
result, the LISA SNR is more sensitive to $|h|$ at large horizon (or
PBH) mass than at smaller masses. This effect is even more pronounced
for the BBO, which is optimized for higher frequencies than LISA.

In fig.~\ref{fig:SNR_N2} the parameter $\tilde{\epsilon}_V$ is varied
as well. Both figures show similar trends and the SNR are very similar
for most of the mass range considered. As has been noted in
section~\ref{sec:PS}, the main effect of increasing $\tilde{\epsilon}_V$
is a suppression of the power spectrum at $k \gg k_i$. For most of the
mass range considered the effect is small; however, at the largest
masses, increasing both $|h|$ and $\tilde\epsilon_V$ can make BBO lose
sensitivity, as seen in the right frame. We note that changes in
$\tilde{\epsilon}_V$ require changes in the scale of inflation in
order to keep the power at CMB scales fixed; the two frames of
fig.~\ref{fig:SNR_N2} therefore have slightly different ranges of
$N_-$.

\begin{figure}[ht]
\centering
\includegraphics[width=0.6\textwidth]{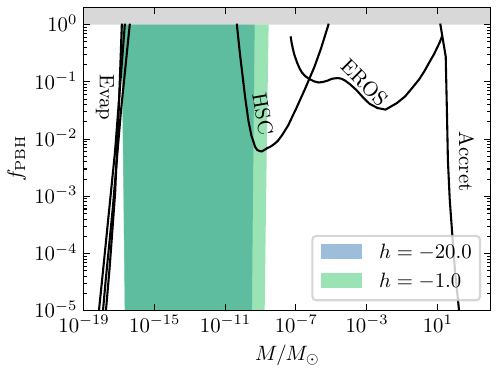}
\caption{If LISA sees no stochastic gravitational wave background, the
  shaded region of PBH parameter space (horizon mass vs. abundance)
  can be excluded.}
\label{fig:LISA_as_f_bound}
\end{figure}

Conversely, one can translate the projected LISA sensitivity curve
into a constraint on PBH production. If no stochastic GW background is
found in LISA, a large window of PBH as DM could be closed. This is
illustrated in fig.~\ref{fig:LISA_as_f_bound}. The shaded region is
the exclusion region from a LISA null result. We have taken
$\tilde{\epsilon}_V=10^{-3}$, and excluded parameters given
$\text{SNR} > 1$ after one year of observation. As the PBH mass
spectrum is not monochromatic in our model, the horizon mass
$M_H (N_-)$ is used to label the $x$-axis in
fig.~\ref{fig:LISA_as_f_bound}. We see that the absence of a signal at
LISA would imply that PBHs in the asteroid mass window contribute
negligibly to the total dark matter density. The value of the
smoothness parameter $h$ determines the upper range of masses that can
be probed in this way; smaller $|h|$ extend the excluded region to the
right, as the higher frequencies of the GW signal are less suppressed
in this case, as noted above. The right edge therefore depends on
details of the second SR phase. In contrast, the left edge of the
shaded area in fig.~\ref{fig:LISA_as_f_bound} mainly depends on the
shape of the power spectrum {\em before} the peak (i.e. at $k <
k_m$). As noted in the discussion of fig.~\ref{fig:compare_DeltaN_h},
here the power spectrum seems to follow a power law, largely
independent of model parameters
\cite{firouzjahiOneloopCorrectionsPower2023,
  piPrimordialBlackHole2023, byrnesSteepestGrowthPower2019}. Hence the
left edge can be considered to be quite robust and universal.
Moreover, recall from fig.~\ref{fig:SNR_N2} that BBO is much more
sensitive than LISA at smaller PBH masses.

\begin{figure}[ht]
\centering
\includegraphics[width=0.6\textwidth]{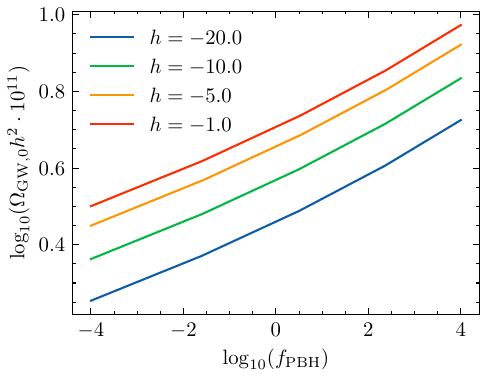}
\caption{Dependence of the strength of the SIGW signal on the PBH
  abundance; for four different values of the smoothness parameter $h$
  (and hence also four different ranges of $\Delta N$).  We have
  assumed $\tilde\epsilon_V = 10^{-3}$ and $N_-=22$ in this plot, but
  the relation between $\Omega_{{\rm GW},0}$ and $f_{\rm PBH}$ depends
  only weakly on these choices. Numerically, to good approximation
  $\Omega_{{\rm GW},0} \propto \left(f_{\rm PBH}\right)^{0.059}$.}
\label{fig:scaling}
\end{figure}

Both edges have very steep slopes, indicating that the PBH abundance
$f_{\rm PBH}$ depends much more sensitively on model parameters than
the strength of the SIGW signal does. This is shown explicitly in
fig.~\ref{fig:scaling}, which shows the dependence of
$\Omega_{{\rm GW},0}$ on $f_{\rm PBH}$; we have taken
$\tilde\epsilon_V = 10^{-4}$ and $N_- = 20$ (so that the PBH mass
distribution peaks just above $10^{-15} \, M_\odot$), but the results
are not sensitive to these choices. We see that changing $f_{\rm PBH}$
by eight orders of magnitude changes $\Omega_{{\rm GW},0}$ by only a
factor of $\sim 3$. This may be surprising at first sight, since the
former depends on $\sim \expval{\zeta^2}$ while the latter depends on 
$\sim \expval{\zeta^4}$,
see eq.~(\ref{eq:SIGW-spec}). However, only very large overdensities,
far out on the tail of the distribution, collapse into PBHs, see
e.g. eq.~(\ref{eq:PS-formula}); the probability for such large
perturbations depends very strongly on various model parameters, as we
saw above. No such threshold effect exists in the SIGW
signal. Figure~\ref{fig:scaling} also confirms that the SIGW signal
decreases when $|h|$ is increased for fixed PBH abundance, as already
shown in figs.~\ref{fig:SNR_N2}.

Finally, we want to highlight the difference between our model and
some previous studies~\cite{firouzjahiInducedGravitationalWaves2023,
  caiOneSmallStep2022}. Because we constructed a realistic inflaton
potential, which allows us to track inflation from the generation of
perturbations at CMB scales through the USR phase until the end, the
power spectrum at small scales has some scale dependence.  This
changes the GW signals at high frequencies. Had we assumed a (nearly)
scale-invariant power spectrum after USR, the right halves of
figs.~\ref{fig:SNR_N2} would have much higher SNR. As a result, the
potential LISA exclusion region in fig.~\ref{fig:LISA_as_f_bound}
would extend several orders of magnitude to the right.

\section{Conclusion and discussion}
\label{sec:conclusion}

In this work, we consider a USR phase of inflation sandwiched between
two SR phases. We allow up to cubic terms in the potential describing
the second slow-roll epoch. This allows us to follow inflation until
the end, while keeping the freedom to vary the speed of the USR to SR
transition, described by the smoothness parameter $h$. The background
equations, as well as the Mukhanov-Sasaki equations describing
curvature perturbations, are solved numerically; the latter
immediately yield the power spectrum at all relevant scales. Using the
$\delta N$ formalism we found that the second slow-roll phase
significantly affects both the power spectrum and the size of
non-Gaussianities. Since the latter could be $\order{1}$ we faithfully
use the full non-linear relation $\zeta(\zeta_G)$, rather than relying
on some perturbative expansion. Next the PBH abundance is calculated
from the probability distribution of curvature perturbations, via the
compaction function $\mathcal{C}$. Since large perturbations, which
are required for PBH formation, inevitably lead to the production of
gravitational waves at second order, we also calculated the resulting
SIGW signal from the curvature perturbation power spectrum.

Observational constraints imply that PBHs can form all of dark matter
only if their mass falls in the asteroid mass range. SIGWs from
formation of such PBHs can be probed by LISA. In order to investigate
the correlation between the PBH abundance and the GW signal
quantitatively, we computed the signal-to-noise ratio at two future GW
observatories, LISA and BBO. It is found that they can both explore
the available BH mass window with $\text{SNR} > 1$. As a result, the
model can be constrained by future GW experiments, even if PBHs only
contribute a (small) fraction of all DM. In fact, according to our
estimate of the SNR, if LISA fails to find any stochastic GW
background, PBHs cannot contribute meaningfully to the total dark
matter density, unless their mass distribution is much broader
than that generated by a local feature in the inflaton potential,
as in our model.

If the PBH abundance $f_\text{PBH}$ is fixed, which can most easily be
done by an appropriate choice of the duration $\Delta N$ of the USR
epoch as measured in $e$-folds of inflation, the most relevant free
parameters of our phenomenological model are the smoothness parameters
$h$ and the duration $N_-$ of the second slow-roll phase.  The former
changes the shape of the curvature perturbation power spectrum at
large $k$, and hence the SIGW spectrum at large frequencies, as well
as the amount of non-Gaussianity. The latter essentially determines
the masses of the PBHs that are produced (if any). As has been pointed
out before, non-Gaussianity enhances PBH production for fixed power
spectrum, and therefore suppresses the GW signals for fixed PBH
abundance.

There are a few caveats, however. It has been discussed extensively
whether $\mathcal{P}_{\mathcal{R}} \sim 10^{-2}$ at small scales (as
required for PBH formation) could ruin the curvature perturbations
detected at CMB scales via loop corrections. If these effects are
unsuppressed, the power spectrum would have to be quite small at all
length scales below those probed by the CMB, in which case density
perturbations of inflationary origin could not lead to a detectable
PBH abundance \cite{kristianoRulingOutPrimordial2022}. However, this
result is not generally accepted \cite{riottoPrimordialBlackHole2023,
  riottoPrimordialBlackHole2023a}. For example, it has been shown that quartic
interactions in the interaction Hamiltonian, which were ignored in the
original paper, are significant as well
\cite{choudhuryNogoFormationHeavy2023,
  inomataCurvaturePerturbationsProtected2024,
  firouzjahiOneloopCorrectionsPower2023,
  choudhuryGalileonInflationEvades2023,
  tadaCancellationQuantumCorrections2024, francioliniOneLoopRule2024}.

In this work, we use the parameters for PBH collapse from
refs.~\cite{gowNonperturbativeNonGaussianityPrimordial2022,
  kitajimaPrimordialBlackHoles2021}. This choice may not be very
accurate. In a more careful treatment of the formation process, one
has to specify the profile of the curvature perturbation rather than
fixing $r$ in terms of the peak location $k_m$ of the power spectrum
\cite{escrivaPrimordialBlackHoles2022}. Moreover, our choice
$k_m r = 2.74$ has only been shown to work well for a monochromatic
power spectrum \cite{kitajimaPrimordialBlackHoles2021,
  muscoThresholdPrimordialBlack2019}; note, however, that our model
does predict a rather narrow distribution of PBH masses, see
fig.~\ref{fig:f_bounds_h=-20}.  A more accurate treatment could be
based on the solution of the equation(s) derived in
ref.~\cite{muscoThresholdPrimordialBlack2019}. Similarly, we have
assumed a sharp cut-off on the compaction function
$\mathcal{C}_c = 0.587$ when computing the PBH abundance.  A more
accurate treatment should include the shape parameter of the
fluctuation \cite{muscoThresholdPrimordialBlack2019}. See also
ref.~\cite{frosinaInflationaryInterpretationNHz2023} for a more
careful treatment of PBH formation.

However, such a treatment, which is considerably more cumbersome, is
really required only once if the precise values of the parameters
describing the inflaton potential are important (in particular, if
they were known). For example, in our model a quite large change in
the prediction of $f_{\rm PBH}$ could be compensated by a small change
of the duration $\Delta N$ of the USR phase. This does not change the
three main results of our analysis: our model allows the production of PBHs
forming all dark matter anywhere in the allowed asteroid mass
range, while allowing for a complete treatment of inflation from
beginning to end; if PBHs in this mass range contribute significantly
to DM, the resulting stochastic gravitational wave background should
be well above the sensitivity limit of LISA, even though our more
careful treatment strongly suppresses the high-frequency part of the
GW signal; and non-Gaussianity, as controlled by the smoothness
parameter $h$, increases the PBH abundance but reduces the GW signal
for fixed $f_{\rm PBH}$, while keeping it well above the sensitivity
threshold.

The last two statements are based on the LISA sensitivity. It is not
completely clear whether any signal to which LISA is sensitive can
also be detected there. Unlike existing gravitational wave
observatories, LISA is predicted to always detect gravitational waves;
in fact, signals from several astrophysical processes, in particular
from binary systems, should be detected at any given time
\cite{dvorkinSyntheticModelGravitational2016,
  auclairCosmologyLaserInterferometer2023,
  gammalReconstructingPrimordialCurvature2025}. Techniques will have
to be, and are being, developed in order to disentangle these various
contributions to the overall signal. We finally note that if LISA is
able to distinguish between astrophysical and cosmological signals, GW
anisotropies could further help constrain the non-Gaussianity
\cite{bartoloGravitationalWaveAnisotropies2020}.

\paragraph{Acknowledgment}
We thank Yong Xu and Andrew Gow for useful discussions.

\appendix

\section{Evolution of the power spectrum during USR phase}
\label{app:power-evo}

\begin{figure}[ht]
 \centering
 \includegraphics[width=0.7\textwidth]{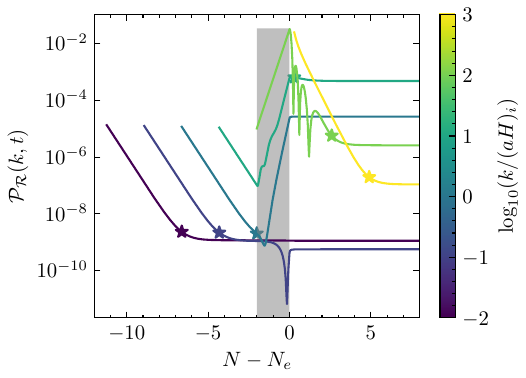}
 \caption{Evolution of the power spectrum, for six different integer
   values of $\log_{10}(k/(aH)_i)$ between $-2$ and $3$.  The stars
   mark the time when a given mode crossed out of the horizon; the
   grey band corresponds to the USR phase, with duration
   $\Delta N = 2$.}
\label{fig:P_evo}
\end{figure}

In this appendix, we explicitly show the evolution of the curvature
perturbation with time, expressed by the number $N$ of $e$-folds of
inflation. Concretely, we solved the Mukhanov-Sasaki equation~\eqref{eq:ms-eq}; the solution gives the full evolution of the power
spectrum $\mathcal{P}(k, t) \propto |u_k/z|^2$.  Figure~\ref{fig:P_evo}
depicts the evolution of modes with
$\log_{10}(k/(aH)_i) = \{-2, -1, 0, 1, 2, 3\}$, where $(aH)_i$ is the
inverse comoving horizon radius at the start of the USR phase. We
choose the following model parameters: $\epsilon_V = 10^{-3}$,
$\beta=10^{-3}$, $N_- = 22$ and $\Delta N=2$. The grey band
corresponds to the USR phase. The perturbations are initialized at
$k/(aH) = 100$ according to the Bunch-Davis condition
\eqref{eq:bunch-davis}.  The star on a given curve marks the point
when of the corresponding mode crossed out of the horizon.

The perturbation with $k/(aH)_i=10^{-2}$ has exited the horizon well
before USR and remains unaffected by it. In contrast, the mode with
$k/(aH)_i = 0.1$ gets reduced in magnitude; this reduction is
responsible for the dip in the power spectrum just before the peak
shown in fig.~\ref{fig:compare_DeltaN_h}. The mode with $k=(aH)_i$,
which exits the horizon right at the start of USR, gets amplified
during USR, despite being (just) outside the horizon; this illustrates
the fact that super-horizon modes need {\em not} be frozen during
USR. An amplification is seen also for $k/(aH)_i = 10$ and $100$;
these modes are inside the horizon during USR. The power in the
$k/(aH)_i=100$ mode oscillates after the USR epoch; this corresponds
to the oscillation (in comoving wave vector $k$) in the power spectrum
after the peak in fig.~\ref{fig:compare_DeltaN_h}. Modes that exit the
horizon long after the USR phase are again unaffected by the USR
phase, as illustrated by the for $k/(aH)_i=10^{3}$.

\section{Variances}
\label{sec:app_var}

In the main text, we have introduced the variances in
eqs.~\eqref{eq:variances}. Here we want to show how they come
about. We assume radiation domination, so that the equation of state
parameter $\omega=1/3$ at PBH formation.

The linear compaction function can be defined via the integral
\begin{align} \label{math:C_l_laplace_zeta}
  \begin{split}
    \mathcal{C}_l(r) &= - \frac{4}{9} r^2 \int_{|\vec{y} \leq r|}
    \dd[3]{\vec{y}} \nabla^2 \zeta(\vec{y}) W_r(\vec{y}) \\
    &= - \frac{4}{3 r} \int_0^{r} \dd{\tilde{r}} \tilde{r}^2 \left(
      \zeta_0''(\tilde{r}) + \frac{2}{\tilde{r}} \zeta_0'(\tilde{r})
    \right)\,,
    \end{split}
\end{align}
where the primes denote derivatives with respect to the argument $\tilde r$,
and the window function is defined by
\begin{equation}	\label{math:window_theta}
W_r(\vec{y}) = \frac{3 }{4\pi r^3}\Theta(r - |\vec{y}|)\,,
\end{equation}
defining the integration region to be a sphere, whose center we have
defined to be origin of the coordinate frame. In the second line of
eq.~(\ref{math:C_l_laplace_zeta}) we have used a multipole expansion of
$\zeta(\vec{y})$ \cite{kawasakiEffectNonlinearityDensity2019}. The
angular integrals then remove all contributions except for the
monopole, which we denote with the subscript $0$ (since
$l=0$). Partial integration then gives
\begin{equation}	\label{math:def_C_l}
    \mathcal{C}_l(r) = - \frac{4}{3} r \zeta_0'(r)\, ,
\end{equation}
which coincides with eq.~(\ref{eq:linear_compaction}) in the main text.

The Fourier transform  of the window function \eqref{math:window_theta} is
\begin{equation} \label{eq:Fourier-W}
W_r(k) = 3 \frac{\sin(kr) - kr \cos(kr)}{(kr)^3} 
     = -\frac{3}{kr} j_1(kr)\,.
\end{equation}
Also writing \eqref{math:C_l_laplace_zeta} in Fourier space, we have
\begin{equation} 
    \mathcal{C}_l (k) = \frac{4}{9} (kr)^2 W_r(k) \zeta_0(k)\,,
\end{equation}
which leads to the often used linear relation between density contrast
and curvature perturbation in eq.~\eqref{eq:delta-zeta-linear}
\cite{escrivaPrimordialBlackHoles2022}. Taking the Gaussian part of
both sides, we have ($X=r \zeta'_{0G}$)
\cite{youngPrimordialBlackHole2019}
\begin{align}    \label{eq:sigma_xx_deriv}
\begin{split}
  \Sigma_{XX} &= \left( \frac{3}{4} \right)^2 \expval{\mathcal{C}_{l, G}
    \mathcal{C}_{l, G}} \\
  &= \frac{1}{9} \int \dd{\ln(k)} (kr)^4 W_r(k)^2 \mathcal{P}_{\cal{R}}(k)  \\
  &= \int \dd{\ln(k)} (kr)^2 j_1(kr)^2 \mathcal{P}_\mathcal{R}(k)\,,
    \end{split}
\end{align}
with power $\mathcal{P}_\mathcal{R} \equiv \mathcal{P}_\zeta$ defined
in eq.~(\ref{eq:PS-u}).

The other random variable $Y = \zeta_{0G}(r)$ can also be expressed as
a smoothing of the curvature perturbation, defined via integration
over the surface of a sphere \cite{youngPrimordialBlackHole2019}:
\begin{equation}
\zeta_{0G}(r) = \int \dd[3]{\vec{y}} \zeta_G(\vec{y}) W_r^s(\vec{y})\,,
\end{equation}
with the spherical-shell function
\begin{equation}
W_r^s (\vec{y}) = \frac{1}{4\pi r^2} \delta(|\vec{y}|-r)\,.
\end{equation}
Once again, only the monopole contribution survives the spherically
symmetric angular integral. The Fourier transform of the
spherical-shell function is nothing but the spherical Bessel function,
\begin{equation}
W_r^s(k) = \frac{\sin(kr)}{kr} = j_0(kr)\,.
\end{equation}
By writing $\zeta_{0G}$ in Fourier space we obtain analogously to
eq.~\eqref{eq:sigma_xx_deriv}:
\begin{equation}
\Sigma_{YY} = \int \dd{\ln(k)} j_0^2(kr) \mathcal{P}_\zeta(k)\,.
\end{equation}
Finally, $\Sigma_{XY}$ is simply the cross correlation between
$C_{l, G}$ and $\zeta_{0G}$, and it leads to the expression in
eq.~\eqref{eq:variances}.

We can understand the variances intuitively as follows. The compaction
function is related to the volume averaged density contrast
(especially at $r=r_m$, where the compaction function is maximal);
this averaging is encoded in the window function $W_r(\vec{x})$. On
the other hand, we care about $\zeta_{0G}$ at a certain $r$ and the $j_0$
comes from the Fourier transform of Dirac delta
\cite{youngPeaksPrimordialBlack2022,
  ferrantePrimordialNongaussianityAll2023,
  ferrantePrimordialNongaussianityAll2023}. Note that a linear transfer
function should for full generality be included in the integral
\cite{ferrantePrimordialNongaussianityAll2023}.

Evaluating the variances accurately is not entirely trivial; the Bessel
functions are extremely oscillatory. Our treatment is as follows: the
zeros of the spherical Bessel functions can be found. This can be done
very naively, since we find that setting the integral upper limit
$kr \sim 10^3$ is sufficient already. Then the interpolated primordial
curvature power spectrum can be used in the integral, which is carried
out by a quadrature method. The zeros of the Bessel function are used
as breakpoints of the quadrature method.

% \clearpage
\bibliographystyle{JHEP}
\bibliography{refs.bib}

\end{document}